\documentclass[twocolumn,amsmath,amssymb]{revtex4}

\usepackage{graphicx}
\usepackage{dcolumn}
\usepackage{bm}

\newcommand{\be}{\begin{equation}}
\newcommand{\ee}{\end{equation}}
\newcommand{\bfig}{\begin{figure}}
\newcommand{\efig}{\end{figure}}

\begin{document}      
\title{Mechanical   response  of  a   self-avoiding   membrane:  fold
collisions and the birth of conical singularities} \author{Paula Mellado}
\altaffiliation[Corresponding author e-mail:]{pmellado@seas.harvard.edu}
\altaffiliation[Current address:]{ School of Engineering and Applied Sciences, Harvard University, Cambridge, Massachusetts 02138, USA}
\affiliation{Department  of  Physics   and  Astronomy,  Johns  Hopkins
University, 3400  North Charles St., Baltimore,  Maryland 21218, USA.}
\author{Shengfeng   Cheng}  
\altaffiliation[Current address:]{ Sandia National Laboratories, Albuquerque, New Mexico 87185, USA.} 
\affiliation{Department  of   Physics  and
Astronomy,   Johns  Hopkins  University,   3400  North   Charles  St.,
Baltimore,     Maryland    21218,     USA.}    
\author{Andres    Concha}
\altaffiliation[Current address:]{ School of Engineering and Applied Sciences, Harvard University, Cambridge, Massachusetts 02138, USA}\affiliation{Department  of  Physics   and  Astronomy,  Johns  Hopkins
University, 3400 North Charles St., Baltimore, Maryland 21218, USA.}

\begin{abstract}
An elastic  membrane that is forced  to reside in  a container smaller
than its natural size will  deform and, upon further volume reduction,
eventually  crumple.   The  crumpled  state is  characterized  by  the
localization  of  energy  in  a  complex network  of  highly  deformed
crescent-like regions  joined by  line ridges. In this article  we study,  
through a
combination  of  experiments,   numerical  simulations,  and  analytic
approach, the emergence of localized  regions of high stretching when a 
self-avoiding membrane is subject to a severe geometrical constraint. Based
on  our experimental  observations and  numerical results,  we suggest
that at  moderate packing fraction, inter-layer  interactions produce a
response  equivalent to that of  a thicker  membrane that  has the
shape  of the  deformed one.  We  find that  new conical  dislocations, 
coined satellite d-cones, appear  as the deformed membrane further compactifies. When these {\em{satellite
d-cones}} are born, a substantial relaxation of the mechanical response of 
the membrane is observed. Evidence is found
that friction plays a key role in stabilizing the folded structures.
\end{abstract}
 \maketitle
\section{Introduction}

\begin{figure}
\includegraphics[scale=0.35]{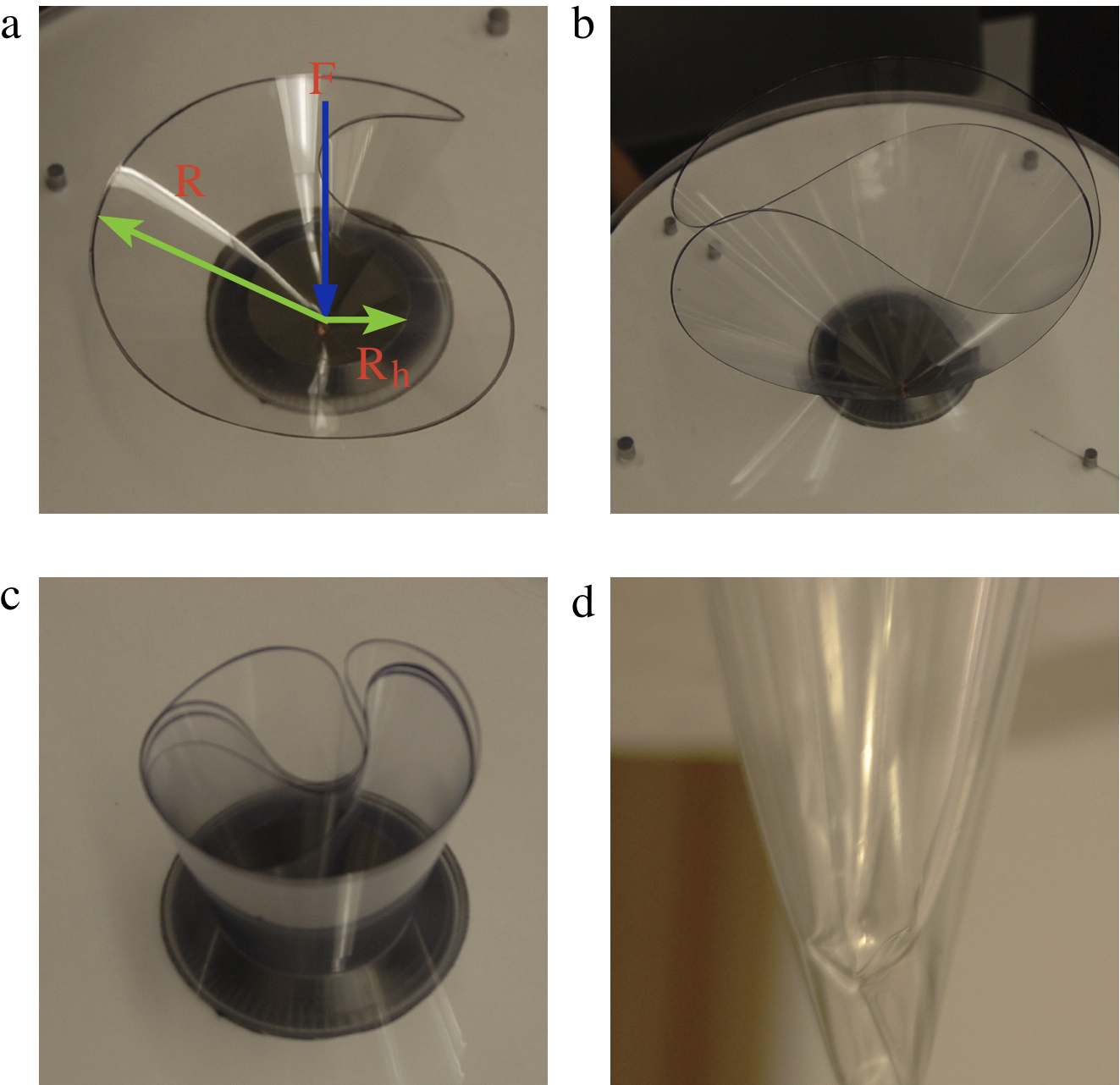}
\caption{\label{F1}(Color  online) Typical  conformations  that appear
 when  pulling  an  almost  inextensible circular  membrane  by  its
 center through a hole. (a), Low load regime in which  a d-cone shape is
 observed. This is an intermediate state of the first regime. (b), Opposite sides  of  the  initially flat  circle
 touch and  self avoidance  becomes crucial. This event marks the onset of the second regime. (c),  and (d),
 are lateral and top  views of high-load
 regimes. The main signature of the second regime is shown in (c) where the mirror symmetry is broken. The satellite d-cone is seen in (d). The onset of the satellite d-cone breaks the translational symmetry along the conical generators. This event marks the onset of the third regime. In (a) the  relevant 
variables to describe this system
 are shown: $R$  is the radius of the membrane,  $R_{h}$ is the radius
 of the hole and $F$ represents the
 load.}
\end{figure}

\begin{figure}
\includegraphics[scale=0.35]{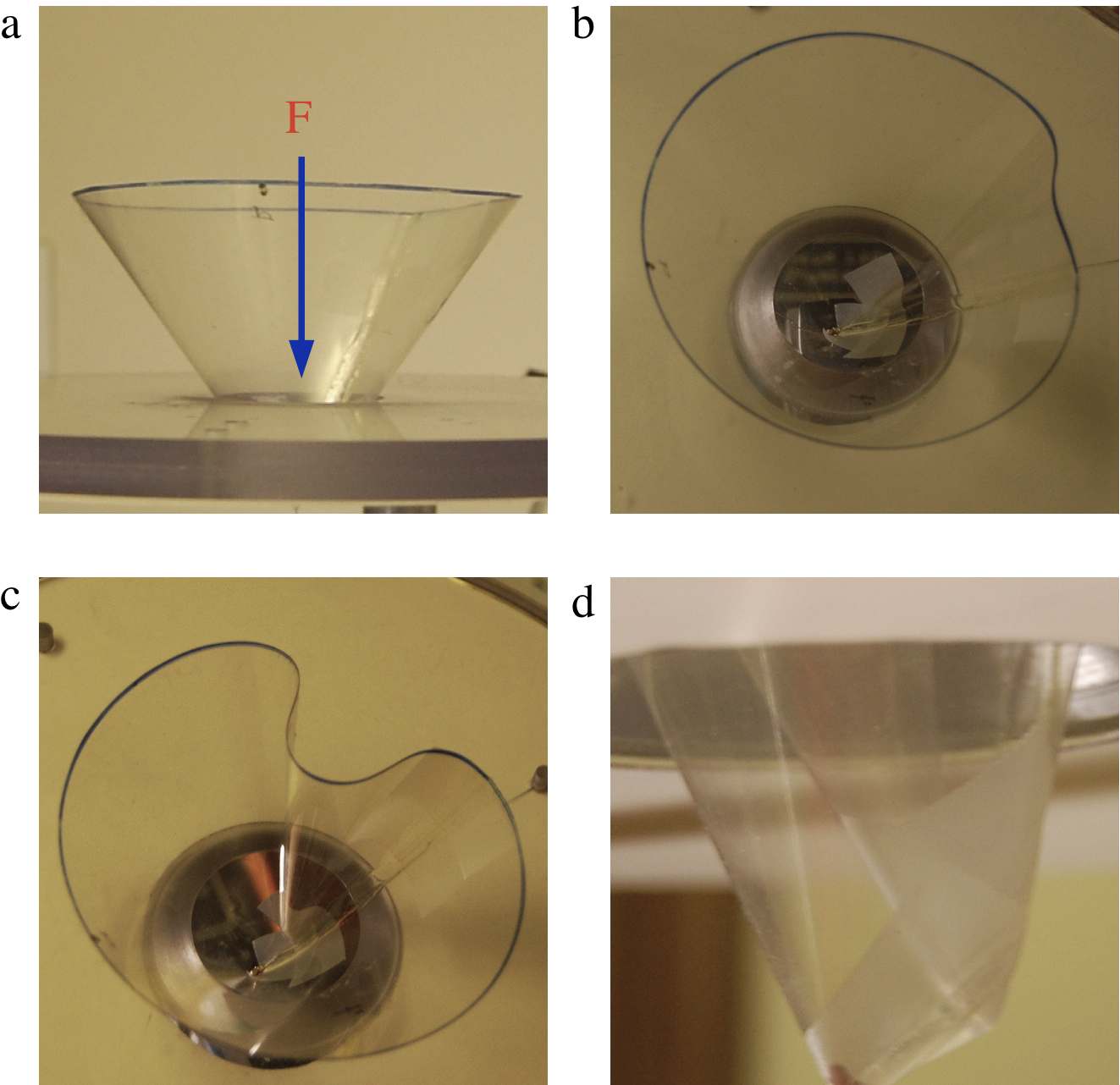}
\caption{\label{F2}(Color  online) Typical  conformations  that appear
 when pulling a conical surface  by its center with a force $F$. (a), Low
 load  regime  in which  the  membrane  is  strained. (b), At  a
 critical  force  a  satellite  d-cone appears.  (c), For  larger
 displacement    in   the   $\hat{z}$    direction   a    hinged   structure
 appears.  (d), One  side of  the  cone touches  the opposite  one,
 resembling the geometry of Fig.\ref{F1}(b).  }
\end{figure}
When  an elastic  membrane is  constrained  into a  small volume,  the
initially   flat   structure   experiences   Euler buckling \cite{landau1995,concha2007}. 
As the  accessible volume decreases, the
membrane goes  through a second  transition generating a  region where
the strain  localizes and  a crescent singularity  is born.  This new
configuration  minimizes  the  total  energy  at the  expense  of  the
emergence of  this pivot point.   Beyond the crescent,  the membrane
forms    a    structure    known    as    developable-cone    (d-cone)
\cite{amar1997,cerda1998,chaieb1998,cerda1999}.  Subsequently, as the
available space  is further  reduced, complex patterns  appear, ridges
that join  these localized objects  forming a network in  which energy
localizes
\cite{lobkovsky1995,lobkovsky1996,kramer1997,didonna2002,vliegenthart2006,wittenRMP2007}.

Besides bending  and stretching,  connectivity and self-avoidance are
also          key           ingredients          in          crumpling
\cite{vliegenthart2006,tchernyshyov2002},    both    determining   the
evolution and the final fate of  the folding process.  On the one hand,
the  connectivity  of a  sheet forces  the  emergence of  the  above-mentioned 
network of highly stretched regions when volume restrictions
are imposed.  On the  other, self-avoidance plays a  key role  in the
mechanical response  of the  crushed membrane since  perfectly elastic
membranes that  incorporate excluded  volume energy are  much stiffer
than phantom  ones \cite{vliegenthart2006}. This  suggests that
it  may be possible  to rationalize  a self-avoiding membrane  as one
effectively thicker than its phantom version.
 
In this paper, we  study a model system that  allows us to make  a natural link
between  the   folding  mechanism  and  the  birth   of  new  crescent
singularities. The experimental setup  consists of a circular membrane
of radius $R$, that under the action of a point-like traction load $F$
directed along  the vertical  direction, is forced  to pass  through a
hole of radius  $R_{h}$. The plane of the hole divides the  space into  two domains
$z>0$ and $z<0$, as shown in Fig.\ref{F1}. We find that the mechanical
response (force  versus displacement plot,  $F$-$z$, see Fig.\ref{F3})
exhibits  a series  of hardening-softening  events as
well as  catastrophic events at  large confinement. A  direct relation
between the emergence of these prominent non-linearities, the frictional
collision of folds, and the  birth of new singularities is reported and
discussed.

The  paper is organized as follows. Section II
is a  description of the experimental  methods and results are reported in Section III.
Section IV is devoted to numerical   simulations as  well as to 
the  analytical results  valid in  certain regimes  of the force versus 
displacement curve.  Finally, Section V  contains  concluding
remarks.
\section{Measurement of the mechanical response}
\begin{figure}
\includegraphics[scale=0.65]{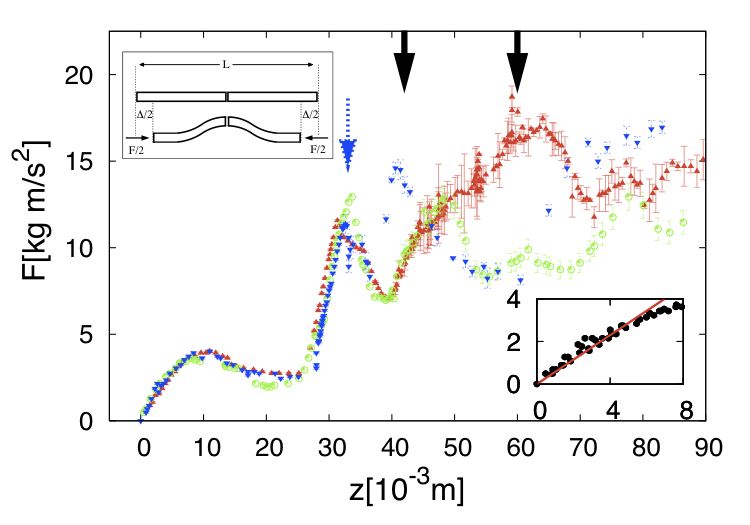}
\caption{\label{F3}(Color  online)  Experimental mechanical  response performed in initially flat membranes:
Force vs.  displacement curves of lubricated (green open circles),  normal (red triangles pointing up) and
sticky (blue triangles pointing down) membranes for increasing confinement. The first regime is realized from $z=0$ to $z \sim 25$ mm, in the three types of membranes. The leftmost (blue) arrow located at $z \sim 33$ mm, shows the first time that two folds collide in the three types of membranes. The  major catastrophic event  corresponds to
the development of a satellite  d-cone (denoted by the middle and rightmost (black) arrows which also denotes the onset of the third regime), which occurs earliest for sticky membranes ($z_{d-cone} \sim 40$ mm) and latest for lubricated ($z_{d-cone} \sim 60$ mm for regular membranes and the satellite d-cone for lubricated membranes is not shown in this plot since it occurs for larger z). The left upper inset shows a schematic representation of the collision of
two folds due to the pushing force $F/2$ at each edge. $\Delta/2$ denotes the displacement of each of the edges of the rod from its initial position along the boundary. The lower inset shows a zoom-in look at the low load regime \cite{cerda1999} along with the best fit obtained using $F=z B/R_{h}^{2}\log(R/r_{c})$ where $B$ is the free parameter reported in the text.}
\end{figure}

The membranes used  in all our  experiments are made  of acetate. Their thickness  is $t\sim 0.10$  mm and their radius
$R=100$ mm. Their bending  stiffness was obtained by measuring the slope at the early stage of the experimental mechanical response curve of the membrane shown in Fig.\ref{F3}, $B= (4.2 \pm 0.1) \times  10^{-4}$  N m (details of the scaling in this stage are presented in Section IV). We performed a tensile test (using the Instron 5544 system) to determine the Young modulus of the membranes, $Y=(3.30 \pm 0.02) \times 10^{9}$ N/$m^{2}$.  The Poisson ratio $\nu \sim 0.35$,  was then computed using the formula $Yt^{3}/12(1 - \nu^{2}) = B$.  The radius of  the hole is fixed at $R_{h}=23.5$ mm.

To pull  a membrane  by its center,  we punched  a small hole at the center 
with a surgical needle and put a thin, stiff wire through
the hole. At one end of the wire we attached a copper ball of diameter
$\sim  2$ mm.  Then, the wire was pulled in order  to produce  a contact
region between the copper ball and the membrane.  Afterward we painted
that  region with  a  small amount  of  silicone glue  to restore  the
membrane  connectivity. We connected  the free  end of  the wire  to a
commercial digital dynamometer (UltraSport)  and a home-made system of
screws  that  allowed  a   controlled  displacement  in  the  vertical
direction  and  the simultaneous  measurement  of  the reaction  force
exerted by the membrane. Special care was taken to make the size of
the perturbed region much smaller than the typical size of the core of
the elastic singularity, which for  the present case was $r_{c}\sim 5$
mm. We pulled the membrane down in the vertical direction $\hat{z}$, at a rate
of $0.015$  mm/s, a factor  $30$ slower than in  a previous work
\cite{boue2006}. This  is an important point since the  mechanical   
response  depends  on  how  fast   the  membrane  is
pulled because of friction.  This  dependence  becomes  stronger when  folds  collide.  In all experiments  the vertical displacement  of the tip  of the
membrane was measured directly with a digital micrometer Mitutoyo. We
observed that each  time the membrane was pulled  a distance $\Delta z$,
it took  at least $2$  seconds for the  sliding process to  stop. For
this reason, every time we pulled the membrane, we waited approximately
$20$ seconds until  the sheets had completely relaxed. This interval was long enough for the reaction force measured by the dynamometer to reach a constant value, confirming that the sheet had reached a stationary state.
As the membrane  progressed along the  vertical direction, it was more difficult to pull it and the force needed to drag it down became larger and larger. But this  process was  non-monotonic: 
every time two folds collide and slide over each other, or a 
catastrophic event such as when a satellite d-cone is born, the force needed to pull and hold the 
membrane drops dramatically, indicating an instability of the packing 
structure of the membrane.

To obtain data with reliable statistics, we repeated each experiment  $10$ times (and each time
we  used  a new  membrane  of  the  same physical  characteristics  as
described above). We studied regular, lubricated, and sticky membranes 
in order to examine  the role  played  by friction. To  obtain
membranes  with  different   friction  coefficients  but  the same mechanical  
properties, we  coated the regular acetate membranes. The reduced friction case was realized by applying a 
thin layer of
graphite powder. We applied the graphite with a clean brush and remove the loose powder.  In the opposite  limit, to increase the  friction we
applied  a thin  film  of $3M$  spray glue, $5$ minutes  before  the
experiment was started. For obtaining as uniform as possible films, we sprayed the glue to the samples from 1 meter apart during 5 seconds.  The change in the thickness of the samples due to the glue coating was measured using a micrometer, being in average of $9$ percent. The change in thickness due to the graphite powder was negligible. The results of the tensile tests (performed with the Instron 5544 system) applied to the coated membranes yielded a change of the Young's modulus due to the coating of $0.3$ percent. The change in the bending stiffness was negligible as well, as apparent from Fig.\ref{F3}. The mechanical response for  these membranes are
shown  in  Fig.\ref{F3}  with  error  bars  deduced  from variations of the experimental data on repeated runs.

To  examine the  deformations that  appear when  an  already compacted
structure  with conical  symmetry is  pulled, we  performed  a similar
experiment with membranes whose initial geometry is conical, Fig.\ref{F2}. Each  cone was cut  out of  a circular
membrane with a deficit angle of $135^{o}$.
We glued  opposite sides of the cut  with epoxy and  used a hair
dryer to release  internal tensions in the material.  We repeated each
experiment  $10$ times  as well for  the same  deficit  angle. We did observe that a d-cone never appeared on the suture line most because along the suture the bending stiffness is larger than in the rest of the structure. The  mechanical
response is shown in Fig.\ref{F4}.  A more detailed description of the
geometry  and  evolution  of  these  deformations  will  be  presented
elsewhere \cite{unpub2010CSM}.

\section{Experimental Results}
\subsection{Compaction of the membrane}
Experimentally we found  three qualitatively different regimes of the 
mechanical response of the membrane. During
the  initial stage the  membrane deforms elastically until  the
pulling force $F$ reaches a  critical value $F_{c}$ at which the first
d-cone  singularity  appears  \cite{cerda1999,cerda1998}. As  we  keep
pulling,  the  single-valued  description  used  in  previous  work
\cite{cerda1998} breaks down. The d-cone structure evolves until it 
touches itself, producing  a contact interaction
between  the two opposite  radial lines  of the  membrane due  to self
avoidance. This new structure is stiffer and, if observed from the top
(See Fig.\ref{F1}b), contains three regions separated by the
contact lines. The first contact between the  two folds marks the end of the
first regime and the birth  of the second and occurs for $z \sim 25$ mm (shown in Fig.\ref{F3} and Fig.\ref{F1}b).  Once in the second regime, the frictional interaction of
two folds at $z \sim 33$ mm, produces a large mechanical response as denoted in Fig.\ref{F3} by the leftmost (blue) arrow.

As the vertical displacement is increased at a constant rate beyond $z \sim 33$ mm, the main
central  fold  shown  in  Fig.\ref{F1}b  is  pushed  against  the
opposite side of the membrane and eventually slops when friction
of the two external folds can not stabilize the system any longer.  After that, a
main spiral develops and breaks the mirror symmetry of a simple d-cone. The break of mirror symmetry is the signature of the second regime (see Fig.\ref{F1}c).
However, in this process the translational symmetry along the conical 
generator is still preserved, particularly outside the core region near the membrane tip. As we keep pulling along $z$, more folds are produced and forced to touch each other, the whole
sheet becomes highly  rigid, and due to self  avoidance, it eventually
shows  an enhanced  mechanical response, a peak signaling the  fact  that the
membrane  is  stuck.  At  this  point,  if  we  increase  the  load further, a
catastrophic  event will  occur: the  formation  of what  we dubbed  a
\emph{satellite d-cone}  that breaks the  local translational symmetry
along the conical generators,  as shown in Fig.\ref{F1}(d) and Fig.\ref{F2}(d). This event
is  characterized  by  a sudden drop in the force-displacement curve,  indicated  by  the middle and rightmost (black)  arrows  in
Fig.\ref{F3} ($z_{\rm{d-scone}}$ depends on friction, $z_{\rm{d-scone}} \sim 40$ mm for sticky membranes $z_{\rm{d-scone}} \sim 60$ for regular membranes) which marks the end of the second regime and the onset of the third one.  The rearrangement of folds allows  the system  to keep
evolving into more compact states as the membrane is being pulled along $z$.  We emphasize that the breaking
of  the  translational symmetry along the conical  generators  makes this  effect  intrinsically  three
dimensional,  and any  attempt to  model  this softening  with a  two
dimensional model is doomed to  fail. Indeed, a study based on a two-dimensional 
description \cite{boue2006} suggested  that at  large compactations
divergent mechanical responses are expected.
 
The emergence of  a satellite d-cone is the  striking signature of the third regime. This is  the simplest realization  of an energy
cascade  proposed  as  the soul  of  crumpling  a long  time ago  and  a
beautiful  example  of  a  nested generation  of  energy  condensation
\cite{cerda1999,wittenRMP2007}.
\begin{figure}
\includegraphics[scale=0.65]{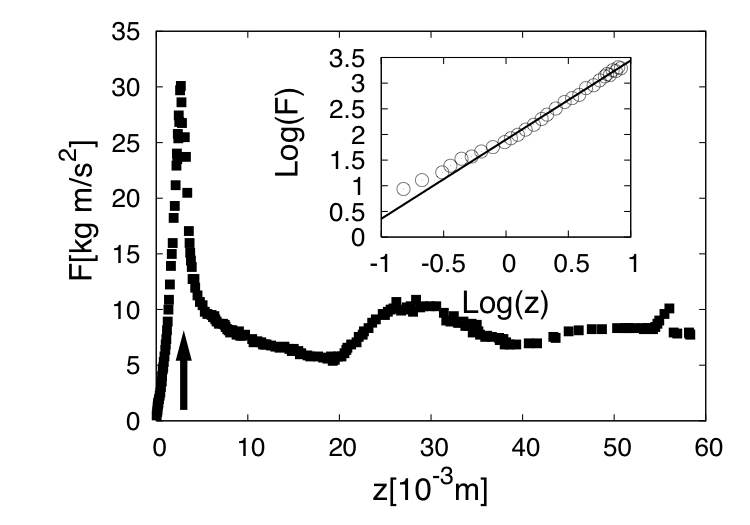}
\caption{\label{F4}Mechanical  response  for initially
conical  surfaces: Force vs. displacement curve.  The
conical  generator length  is  $R=100$  mm and  the  deficit angle  is
$\theta=135^{o}$.  The actual  membrane  and deformation  is shown  in
Fig.\ref{F2}. The arrow indicates the onset of a satellite d-cone. The inset shows a zoom-in look to  the mechanical response
before the first satellite d-cone appears (open circles) along with the best fit (line). In this region $F\sim z^{1.50\pm0.02}$. Error bars are smaller than the size of the data points and thus not shown here.}
\end{figure}
The three curves shown in Fig.\ref{F3} clearly show that friction does
play an important  role in the mechanical response  of our system. 

Friction 
in  two dimensional packing  of wires was  shown to
have  a  strong impact  in  the  packing morphology  \cite{stoop2008},
showing   the so-called   classical,   spiral,   and   plastic
morphologies. However,  in two dimensions  crumpled wires do  not have
the freedom to  release energy by deforming into  another dimension as
in  the case discussed  here. In  two dimensional systems,  a
divergent  mechanical  response  close  to a  non-universal  (material-dependent)  
critical packing  ratio is generically  observed, which for  certain parameters  was
found  to  be  $\phi_{c}\sim   0.46$  \cite{stoop2008}.  Our  case  is
different and does have a  third dimension that  allows new deformation modes. One example is the satellite d-cone observed here. We  further found that  the  vertical
displacement  at  which the  satellite  d-cone  forms depends  on
friction. The middle and rightmost (black) arrows in Fig.\ref{F3} indicate that the 
satellite d-cone appears earlier when the friction is larger. This
can be qualitatively understood
in the following  way: larger friction means it is harder for 
touching layers to slide over each other, which makes the 
membrane effectively more rigid. Therefore, for membranes 
with a larger friction coefficient, the compaction preserving 
the translational symmetry along the conical direction stops 
at smaller vertical displacements. As the displacement 
is further increased, the satellite d-cone appears to 
allow the membrane to further compact.

All our experimental and simulation results indicate that  the satellite
d-cone is the basic deformation mechanism of a conical surface, which is 
effectively the geometry of the flat membrane when it 
is in a compacted state with translational symmetry. To provide 
more evidence of this statement, we performed an
experiment of a membrane whose initial geometry was conical.  This experiment
(see  Figs.\ref{F2}-\ref{F4}) shows   that  for  large
confinement, once  the folded membrane is stuck  at a certain friction  dependent position, the  system  starts to stretch until  it reaches
  a critical force $F_{c}$ at which energy focuses in a localized region. This region is at a certain distance away from the membrane tip and is the core of the satellite d-cone.

The mechanical response of the cone is shown in Fig.\ref{F4}.  
The first  stage is  characterized  by a 
dragging force that grows as a power law $F\sim z^{q}$, where $q=1.50\pm
0.02$, as  obtained from the best  fit of the  $F-z$ curve shown 
in the inset of Fig.\ref{F4}. In this regime
the  overall symmetry  of the  cone remains  unchanged,  therefore the
measured force can only be due to the stretching of the membrane.  After a
critical load  $F_{c}$ has been  reached, or equivalently, after a critical
amount of energy  has been stored in the  stretching mode, the conical
shape becomes unstable and  a satellite  d-cone is  born, as  shown in
Fig.\ref{F2}b and  as indicated by the arrow in Fig.\ref{F4}. The  energy transfer  between
stretching  and bending modes  is reflected  in the  $F$-$z$ curve  as a
sudden  drop  in $F$,  similar to  what  is
observed  in Fig.\ref{F3} for a flat membrane in a compacted 
state. After  the satellite  d-cone is  formed, the
system evolves in a similar way to the evolution of the initially flat membrane
at  intermediate packing  fraction.  In this  regime the $F$-$z$ curve 
 goes down  until opposite  sides of  the cone
touch each other.

\subsection{The role of friction}
Fig.\ref{F3} shows that at small vertical displacements, friction and self-avoidance play no role because different parts of the membrane do not touch yet. In this region, all membranes with different frictional treatments exhibit the same mechanical response and the force-displacement curves collapse. However, we  do not  anticipate any
possible  collapse   of  the  $F$-$z$   curves  beyond  the   first  fold
collision. In this region, the different friction properties of the membranes affect the collision and sliding of folds, leading to friction-dependent mechanical responses, like  the
non-universal  behavior   found  in  the  compactation   of wires in 2D
\cite{stoop2008}.  

Generally, for larger friction 
wider folds are observed before they slip. This  observation allowed  us to
further unveil the role of  friction quantitatively. We
estimated its  contribution by  computing the friction  coefficient in 
two ways.  In the first approach  we analyzed a
fold collision at low packing fraction (which is representative of the
processes happening at more compact configurations). An example is  
shown in  Fig.\ref{F1}b, where  two folds  touch each  other. This
state will last  as long as the forces  in the direction perpendicular
to the contact line do  not overcome the frictional forces.
As the pushing force increases due to increasing confinement, both 
folds become wider and eventually
buckle  as if  they  were a  single  rod, forming  an effective  Euler
elastica  (see schematic representation shown in the inset of  Fig.\ref{F3}). At this  stage the
contact point will work as a joint and the folds will move with respect to
each other: the reaction force will increase until they slip.  We make
use of this description to  estimate the stability of the collision of
two folds in the experiment.  The only force stabilizing the system is
friction, $\mu  \bf{N}$, in  which $\mu$  is the  static coefficient of friction
and $\bf{N}$ is the normal force. By virtually cutting an Euler elastica into two halves, we compute the bending  force needed to  keep the
system together with the virtual work  technique.  Equating these
two forces yields
\begin{eqnarray}
\mu \sim \frac{\pi}{2(1-\nu^{2})}\sqrt{\frac{\Delta}{L}},
\label{mu1}
\end{eqnarray}  
where $L$ is the size of the effective rod, $\nu\sim 1/3$, its Poisson
ratio, and $\Delta$ is the displacement of the edges of the rod from
its  initial  position along  the  boundary.  Experimentally we  found
$\Delta/L\sim   1/15$. Thus $\mu$ can be estimated using Eq.\ref{mu1}. In the second approach, we obtained $\mu$ using a traditional method: An incline and a block that  is free to slide as soon as the
incline   reaches  a  critical   angle  $\theta_{c}$.    The  friction
coefficient in this case is well known to be given by
\begin{eqnarray}
\mu=\tan\left(\theta_{c}\right)
\label{mu2}
\end{eqnarray}
The coefficients of friction obtained by these two independent methods
agree well within experimental error  $\mu=0.45\pm 0.05$. This indicates that
 the set of most  prominent bumps observed in Fig.\ref{F3}, at low
packing fraction  are a consequence of the  frictional collision between
folds and that the height and  position of these events in the $F$-$z$
curve  becomes material  dependent since they are determined by friction.

In our experiments using membranes lubricated with graphite powder, the 
friction is very low and the onset of
the satellite  d-cone occurs for  a larger load  than in the case  of the
regular membranes. In other words, the satellite d-cone occurs later in the $F$-$z$ curve. To study the opposite limit where friction is 
extremely large, we deposited a  thin film of a 3M glue on
top  of the  acetate membrane. This treatment makes the folds 
stick once they touch each other and they remain stuck 
unless the adhesive layer fails. But the failure never occurred 
in our experiments. In this extreme case, the membrane can only
 invaginate from the configuration in which the two main folds shown
in Fig.\ref{F1}b are glued together.  We discovered that the effect of
this strong attractive contact  interaction is to produce a cone like 
the one shown  in Fig.\ref{F1}c earlier than for the regular
membranes.  Consequently, the  onset  of the  satellite d-cone  occurs
earlier in the $F$-$z$ curve (see middle (black) arrow in Fig.\ref{F3}).

\section{Molecular Dynamics Simulations}

\begin{figure*}
  \begin{center}
    \begin{tabular}{cccccc}
      {\bf{1}}  &  {\bf{2}}  &   {\bf{3}}  &  {\bf{4}}  &  {\bf{5}}  &
{\bf{6}}\\  {\bf{a}}  \resizebox{23mm}{!}{\includegraphics{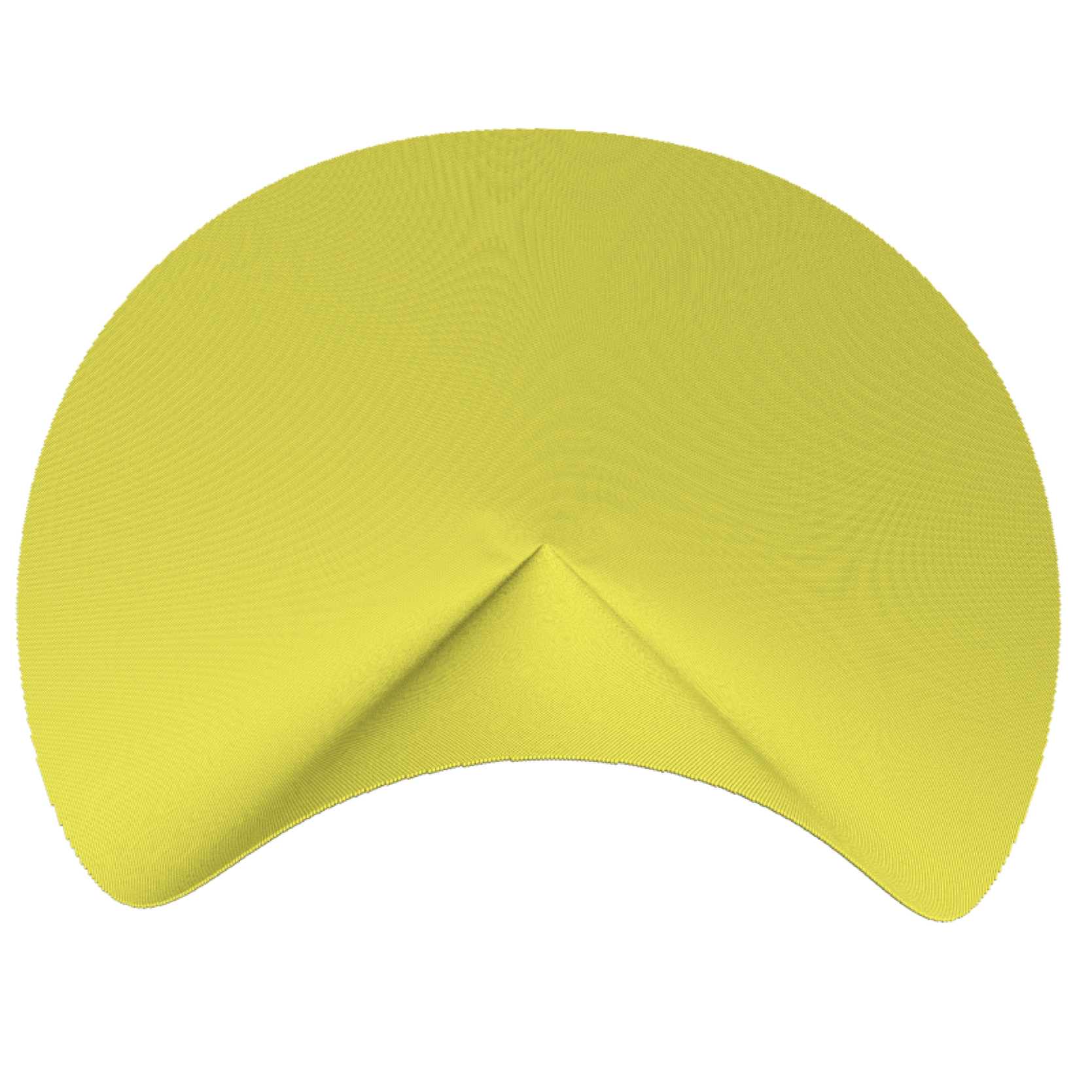}}
&\resizebox{23mm}{!}{\includegraphics{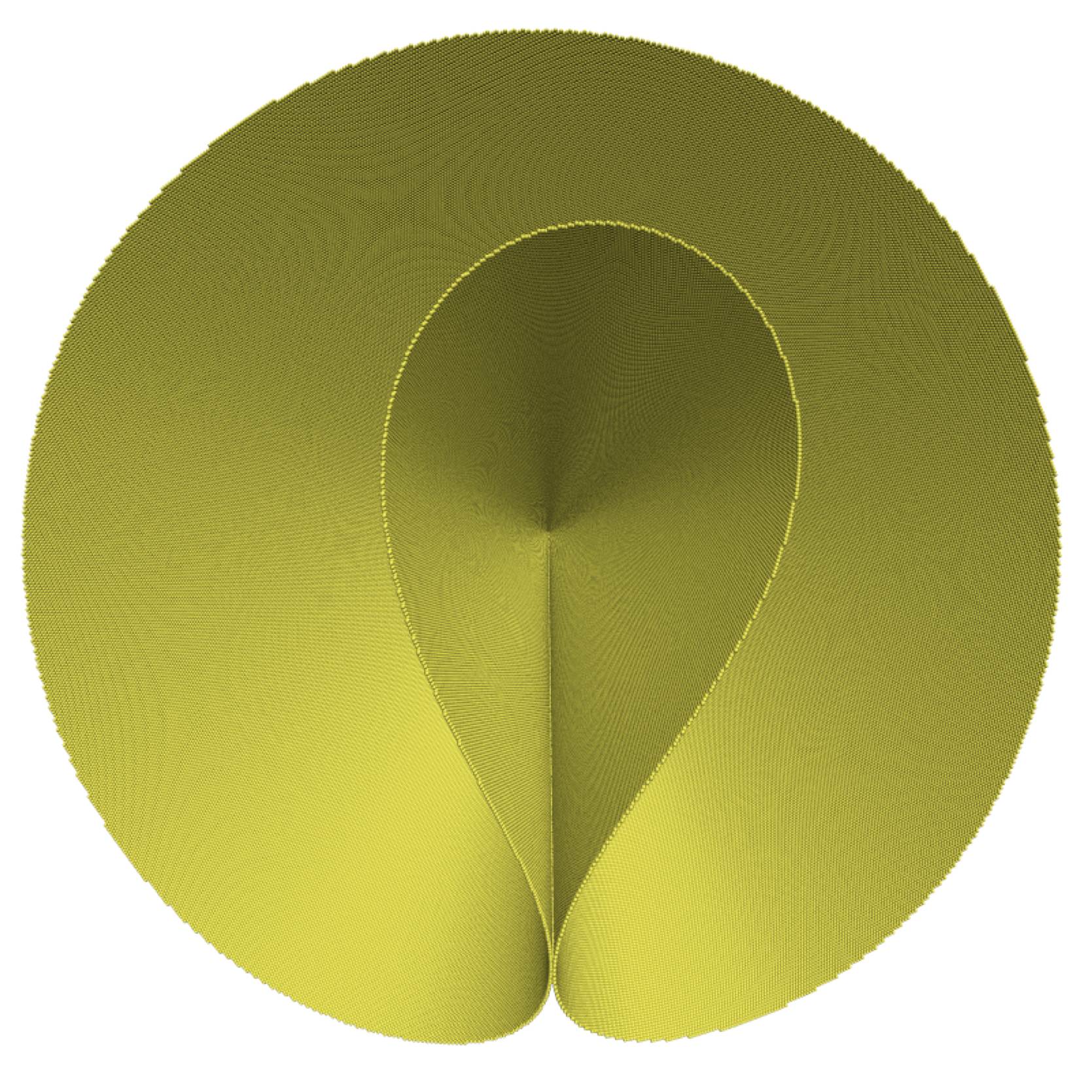}}&\resizebox{23mm}{!}{\includegraphics{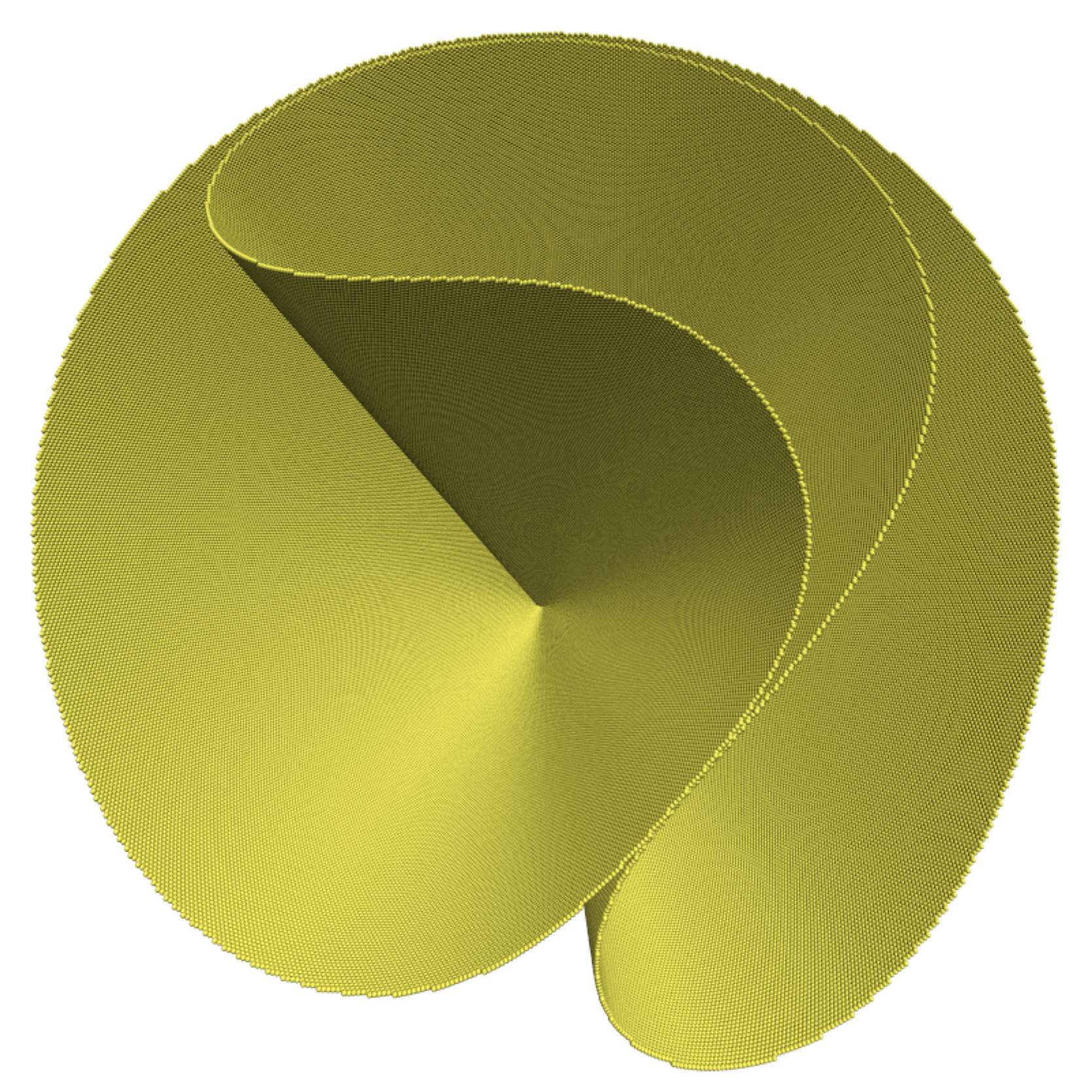}}&
\resizebox{23mm}{!}{\includegraphics{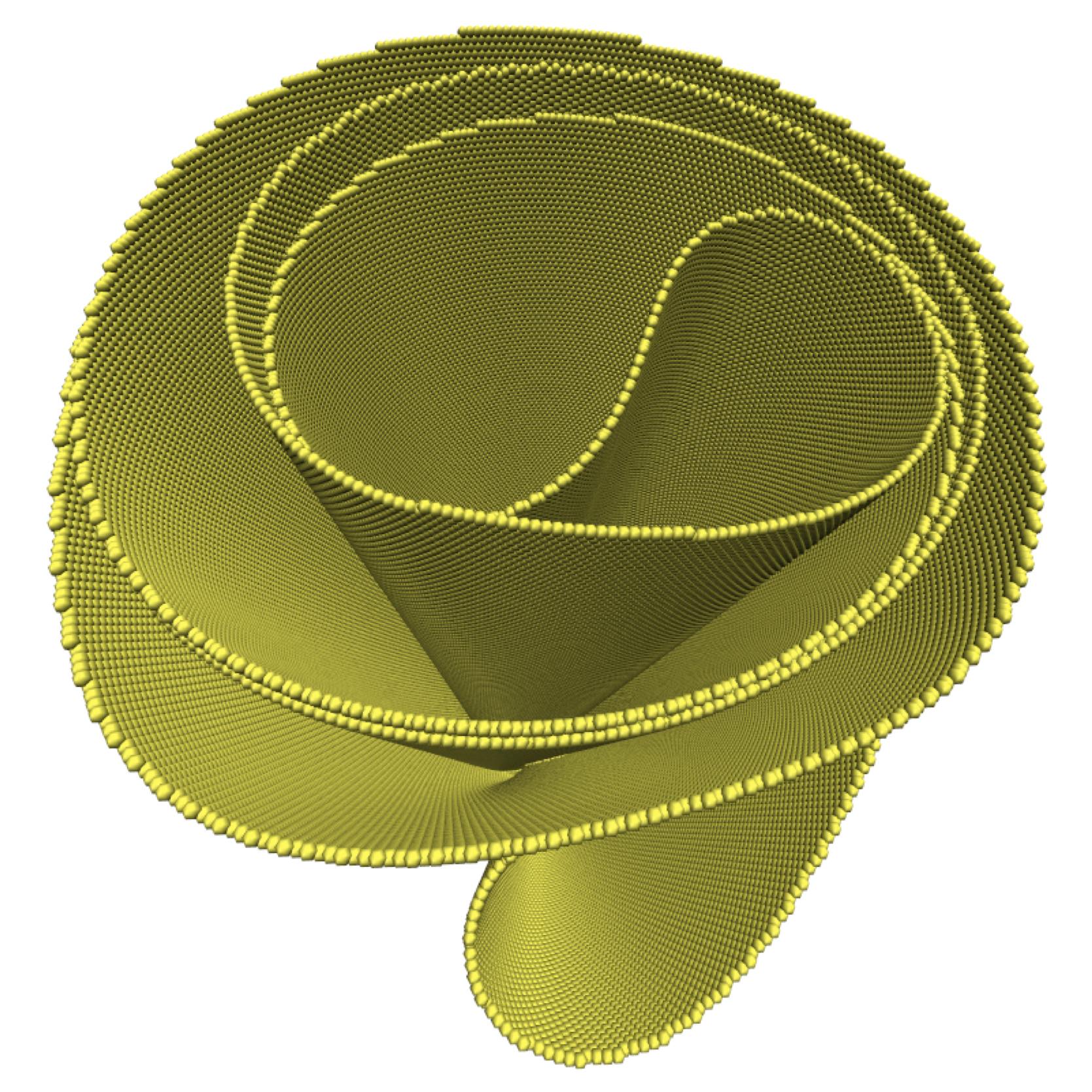}}&\resizebox{23mm}{!}{\includegraphics{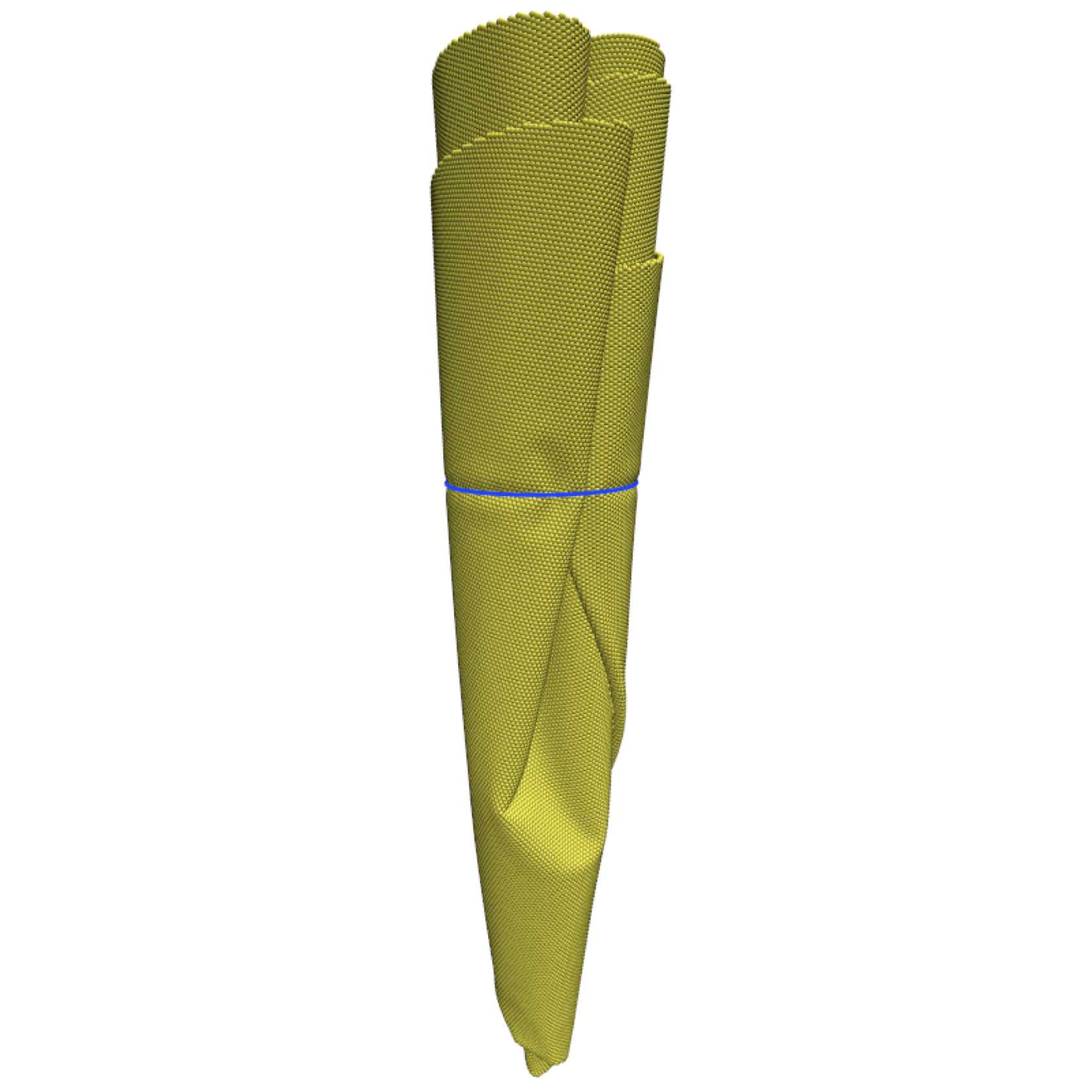}}&
\resizebox{23mm}{!}{\includegraphics{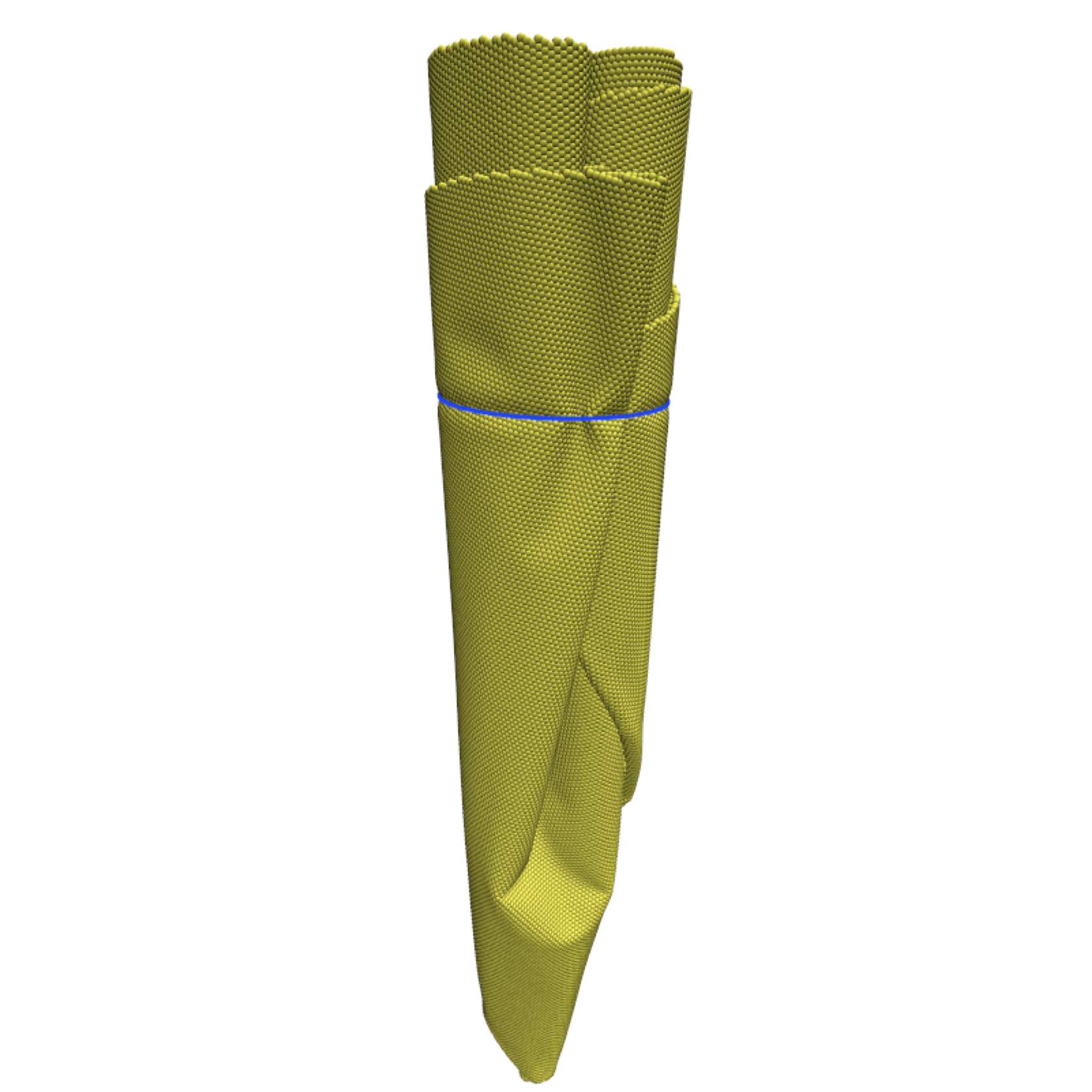}}      \\      {\bf{b}}
\resizebox{23mm}{!}{\includegraphics{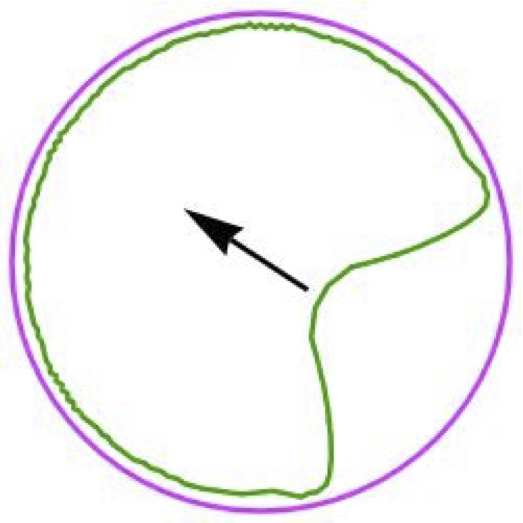}}
&\resizebox{23mm}{!}{\includegraphics{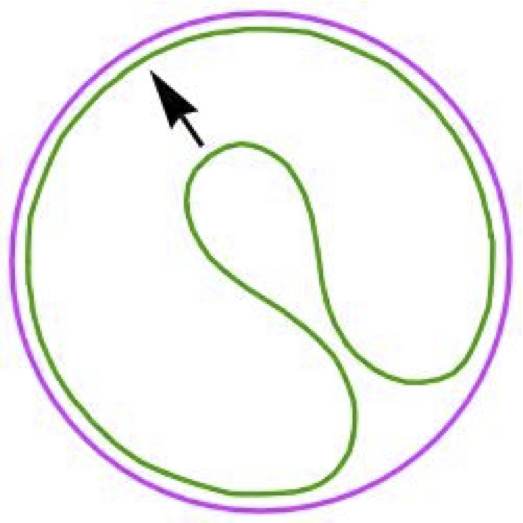}}&\resizebox{23mm}{!}{\includegraphics{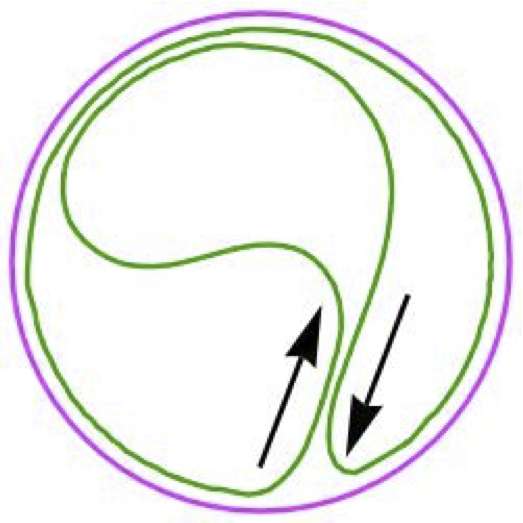}}&
\resizebox{23mm}{!}{\includegraphics{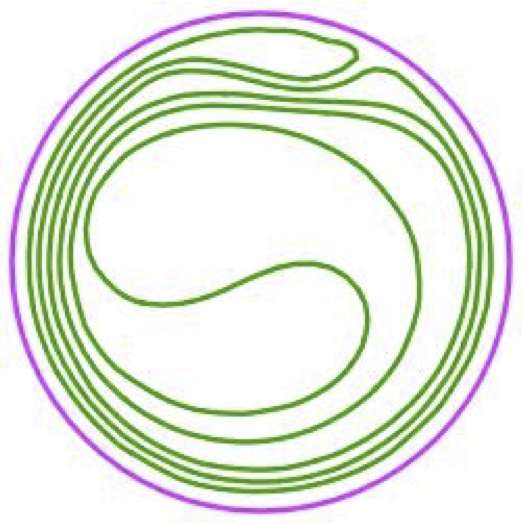}}&\resizebox{23mm}{!}{\includegraphics{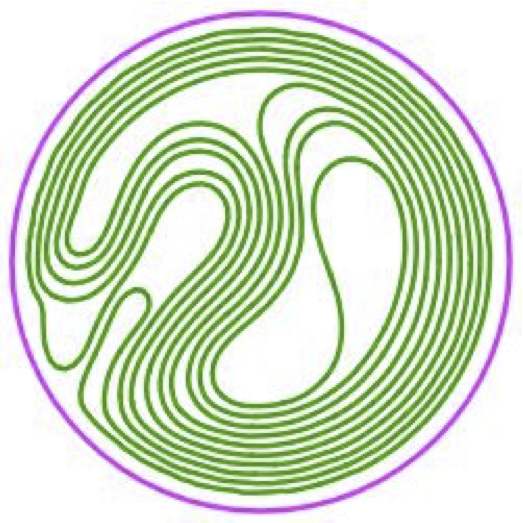}}
&    \resizebox{23mm}{!}{\includegraphics{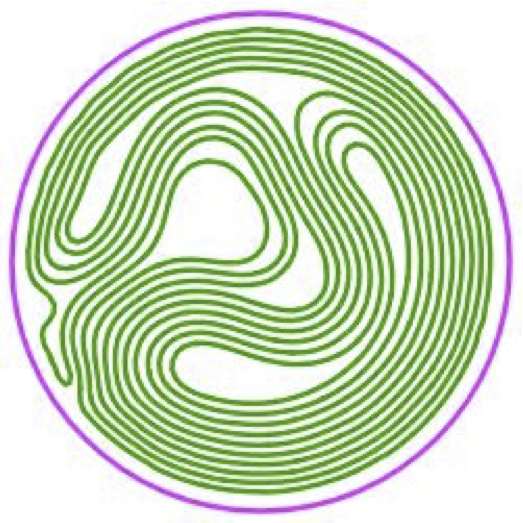}}    \\   {\bf{c}}
\resizebox{23mm}{!}{\includegraphics{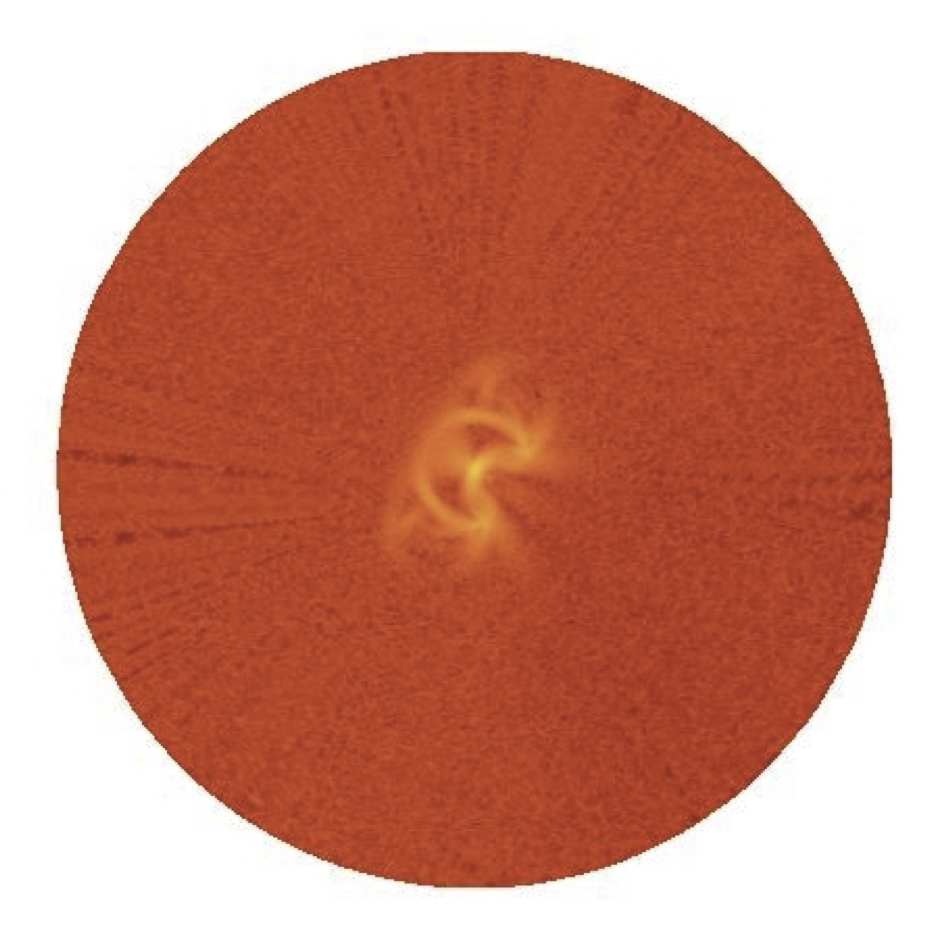}}
&\resizebox{23mm}{!}{\includegraphics{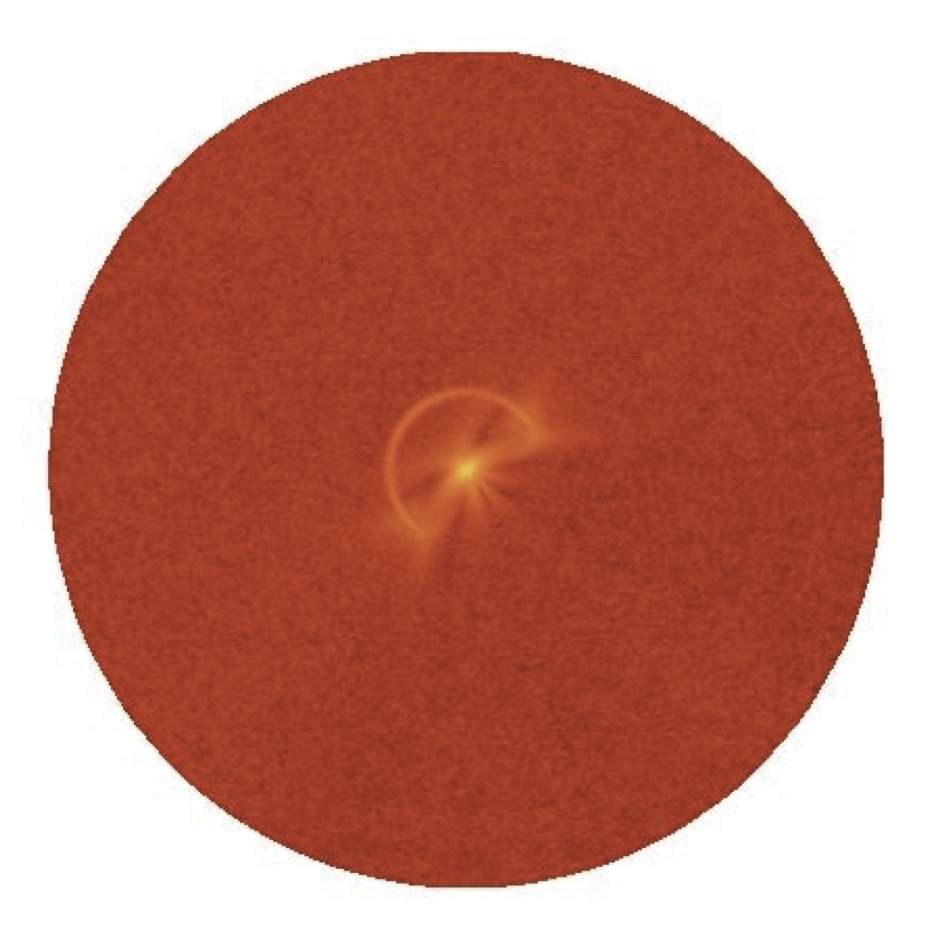}}&\resizebox{23mm}{!}{\includegraphics{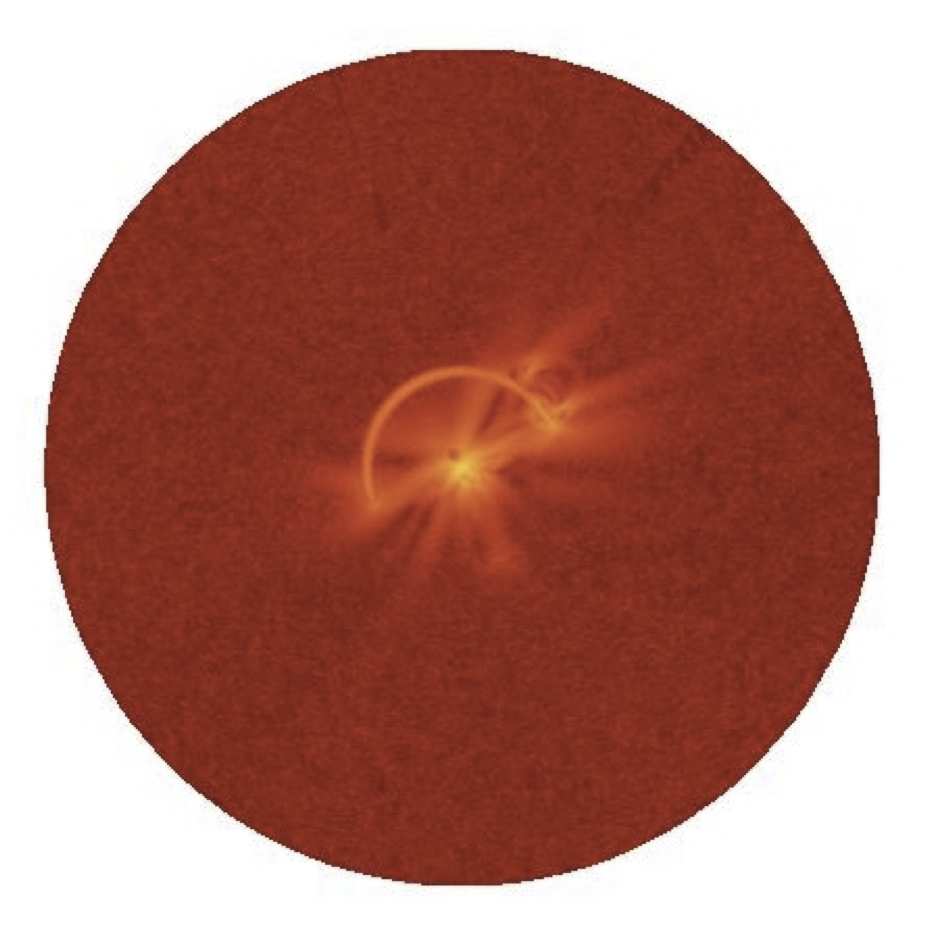}}&
\resizebox{23mm}{!}{\includegraphics{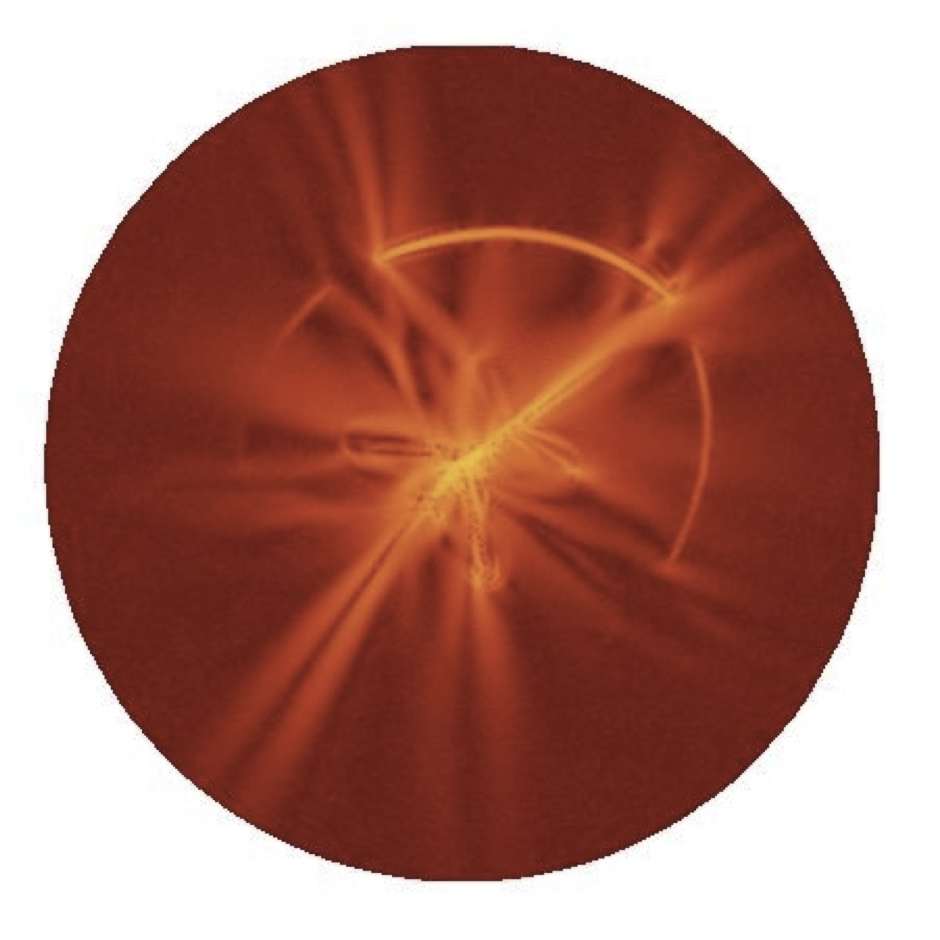}}&\resizebox{23mm}{!}{\includegraphics{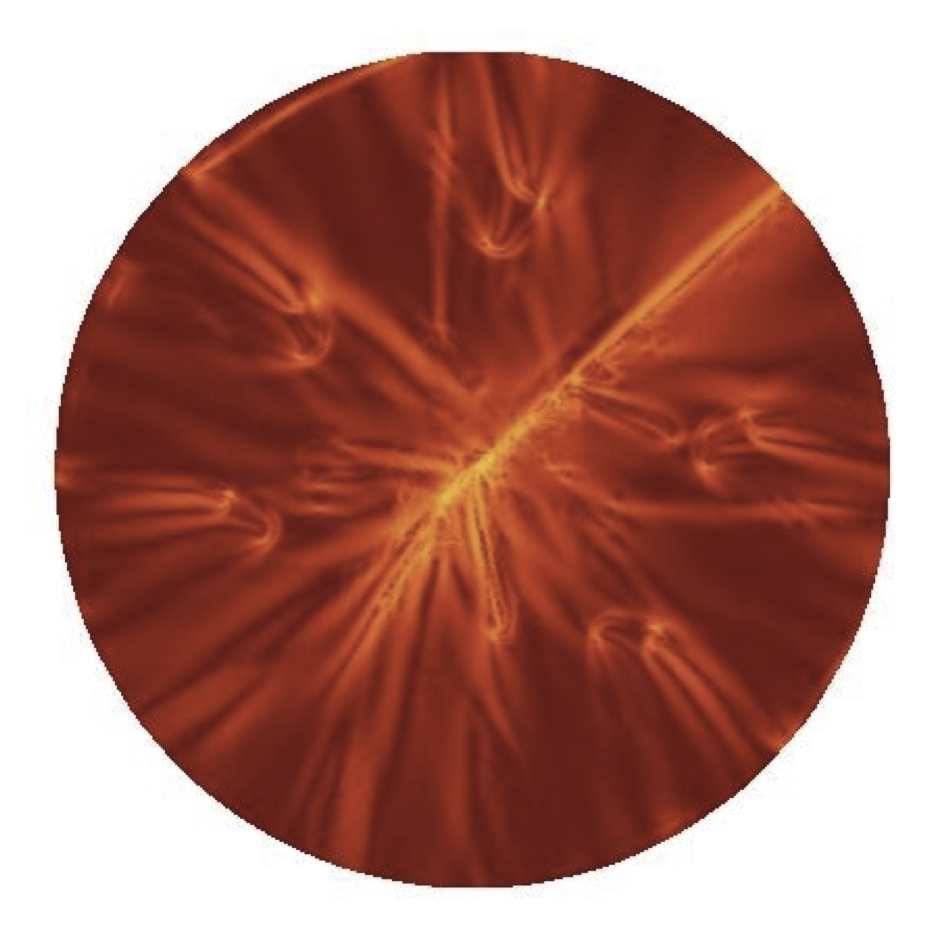}}
&\resizebox{23mm}{!}{\includegraphics{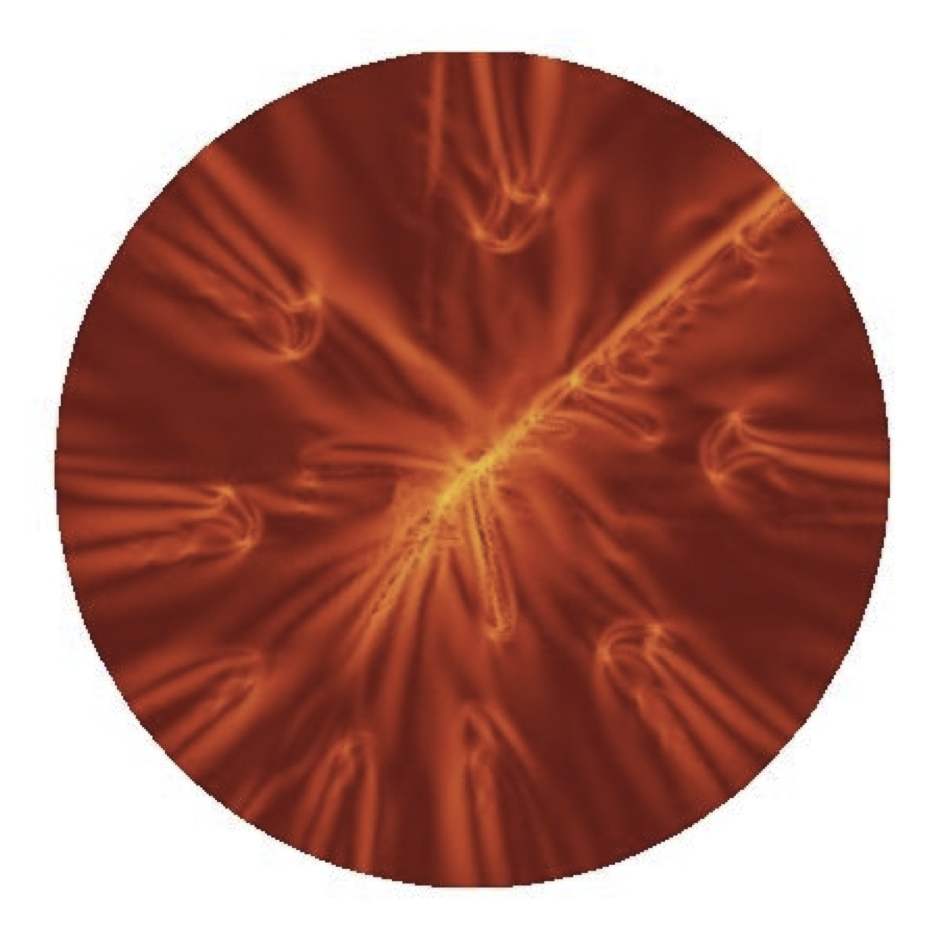}}      \\     {\bf{d}}
\resizebox{23mm}{!}{\includegraphics{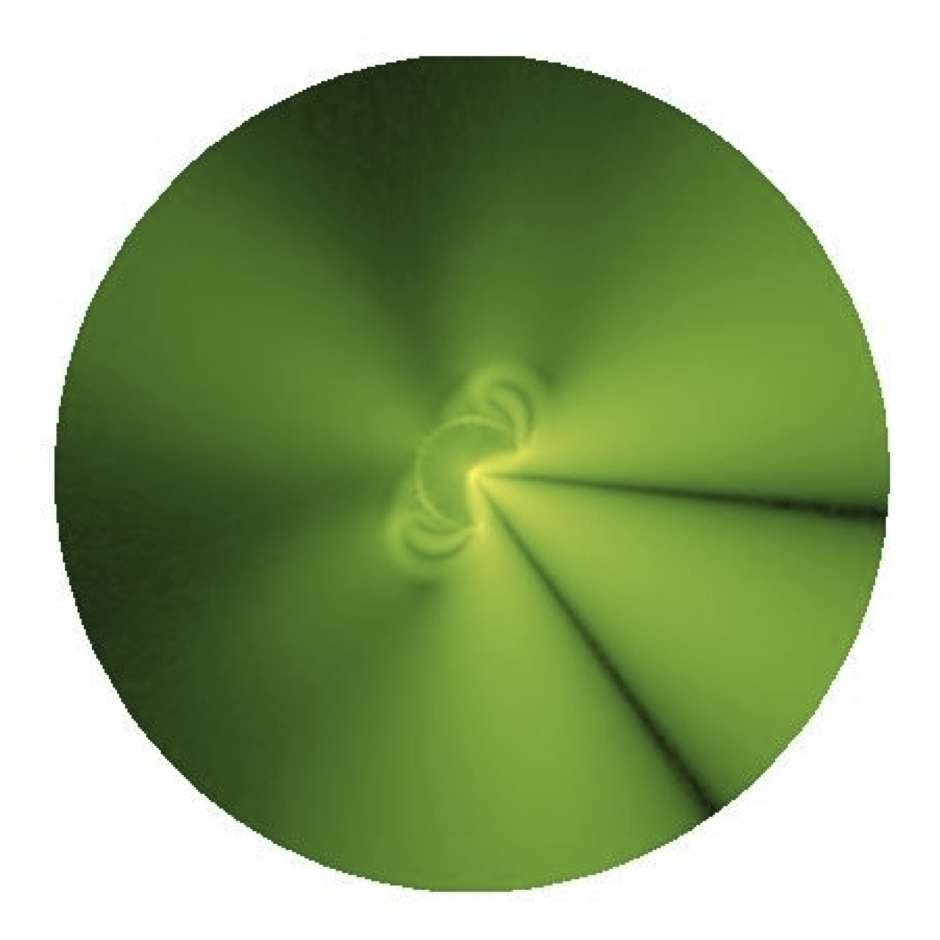}}
&\resizebox{23mm}{!}{\includegraphics{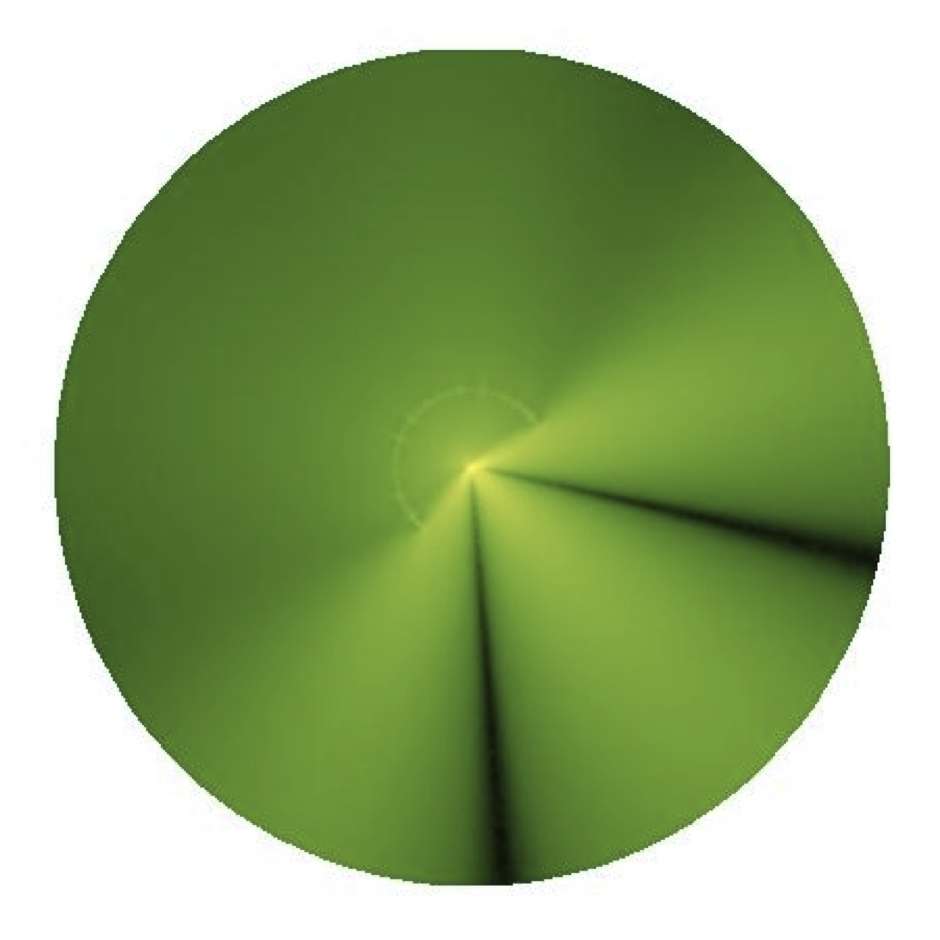}}&\resizebox{23mm}{!}{\includegraphics{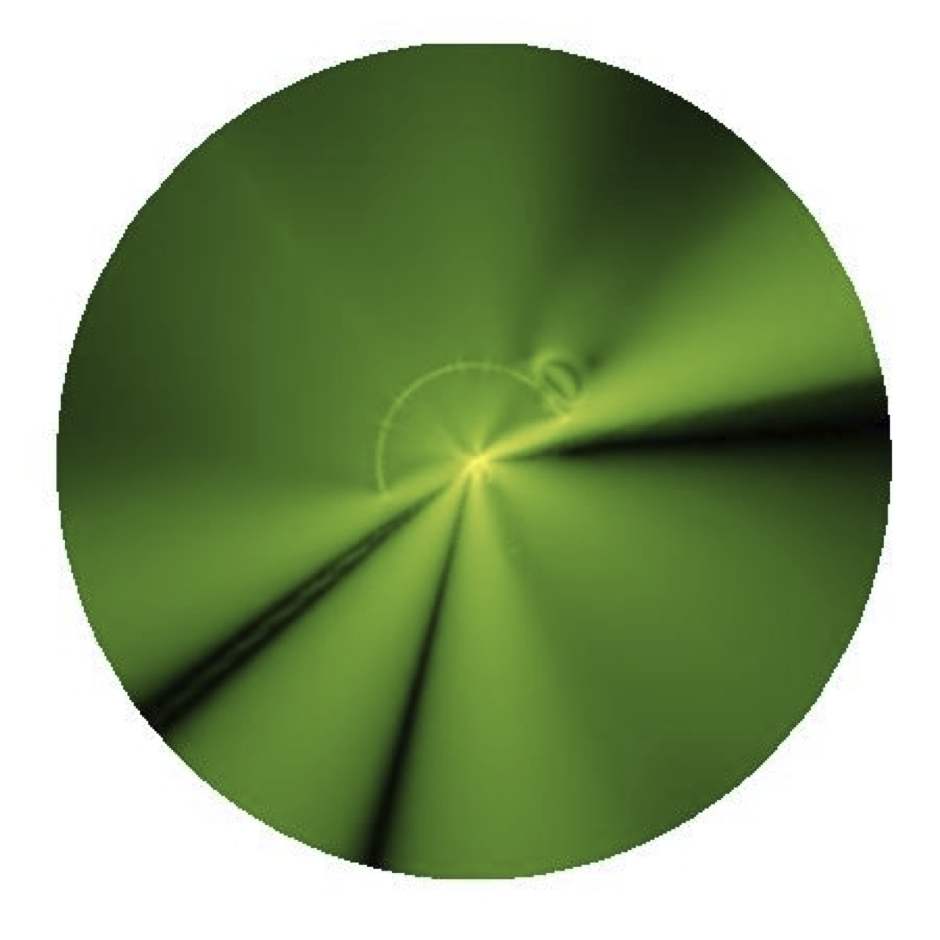}}&
\resizebox{23mm}{!}{\includegraphics{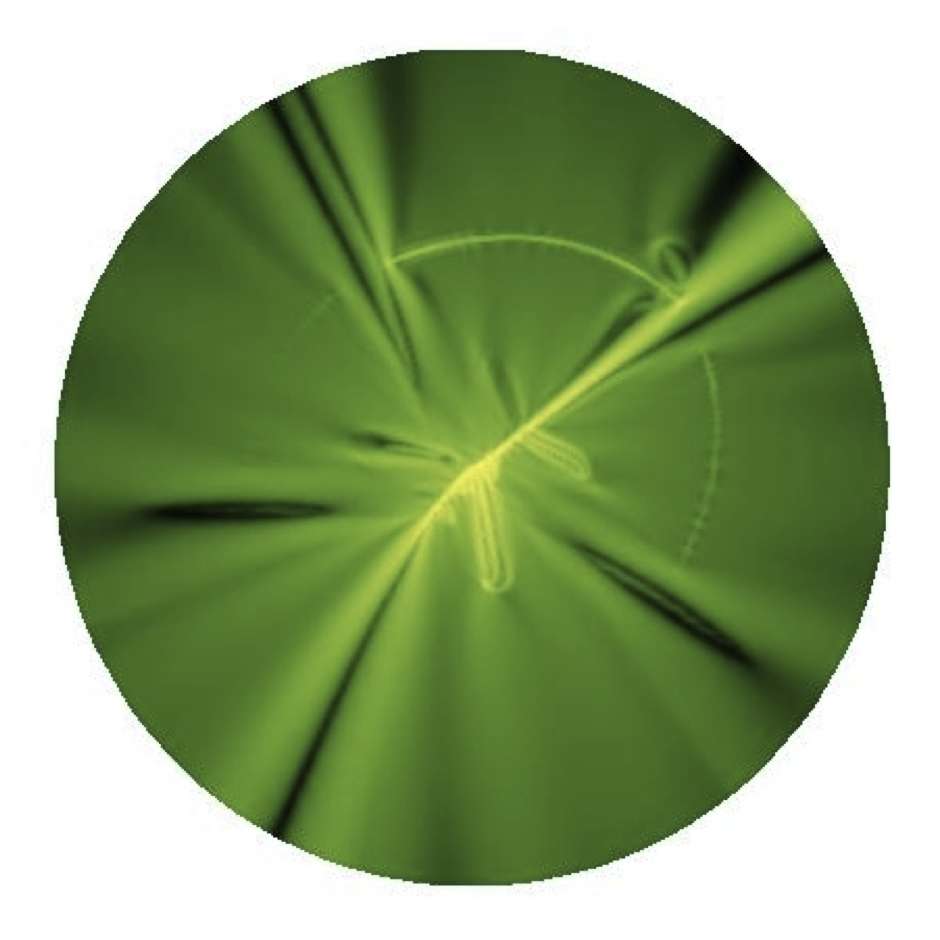}}&\resizebox{23mm}{!}{\includegraphics{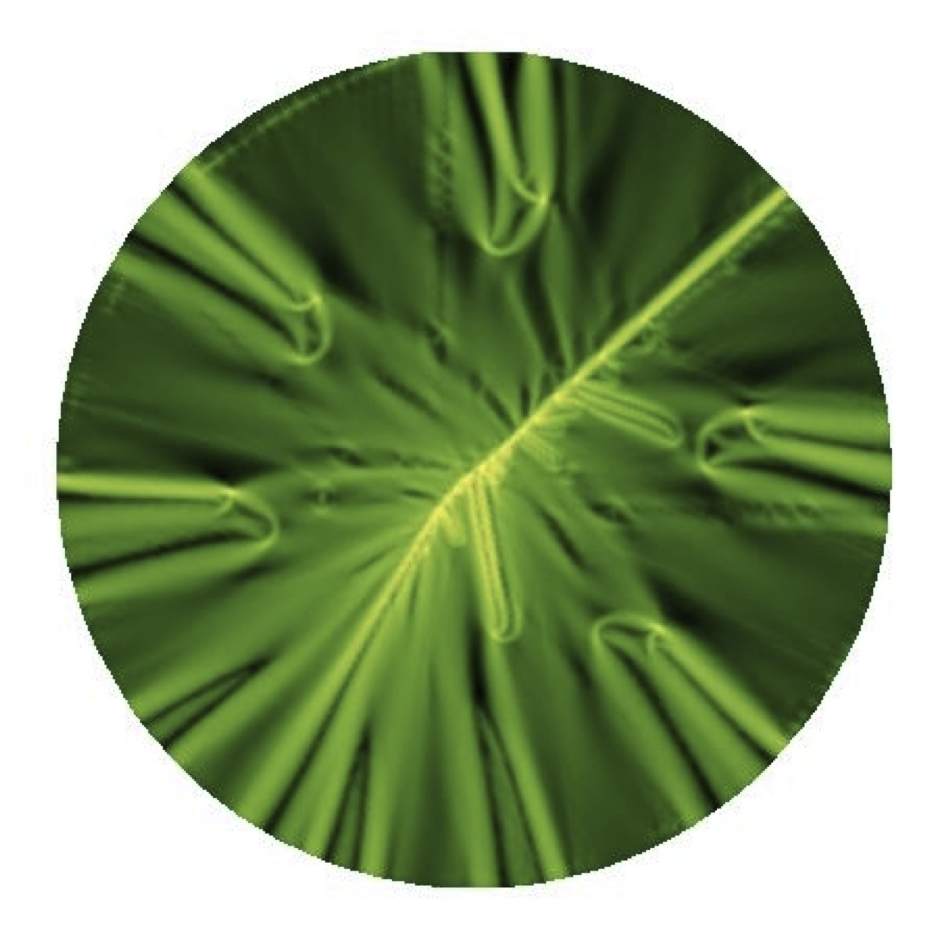}}
&    \resizebox{23mm}{!}{\includegraphics{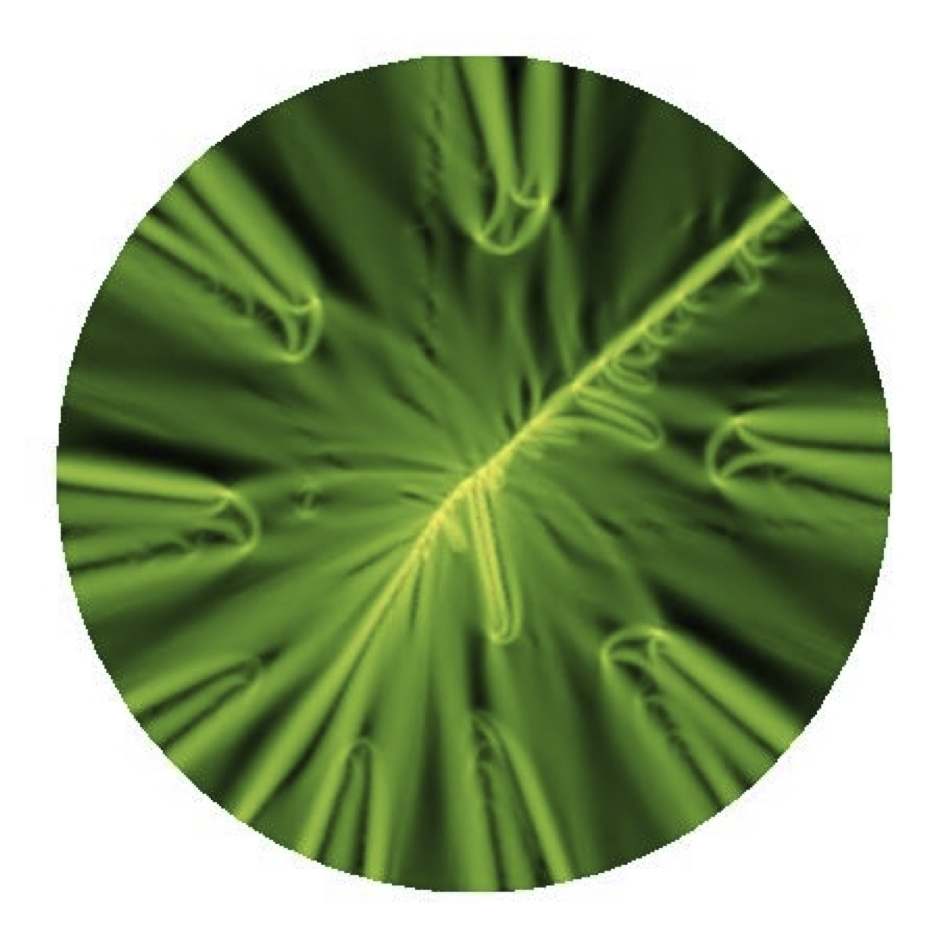}}    \\   {\bf{e}}
\resizebox{23mm}{!}{\includegraphics{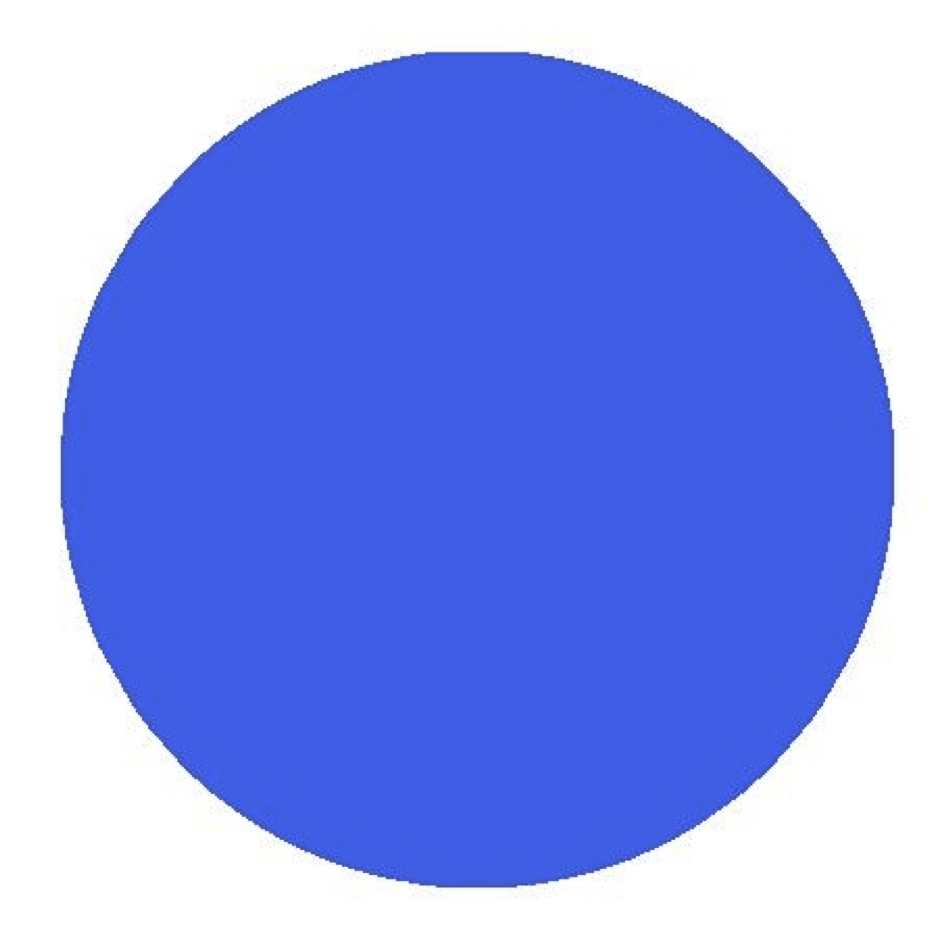}}
&\resizebox{23mm}{!}{\includegraphics{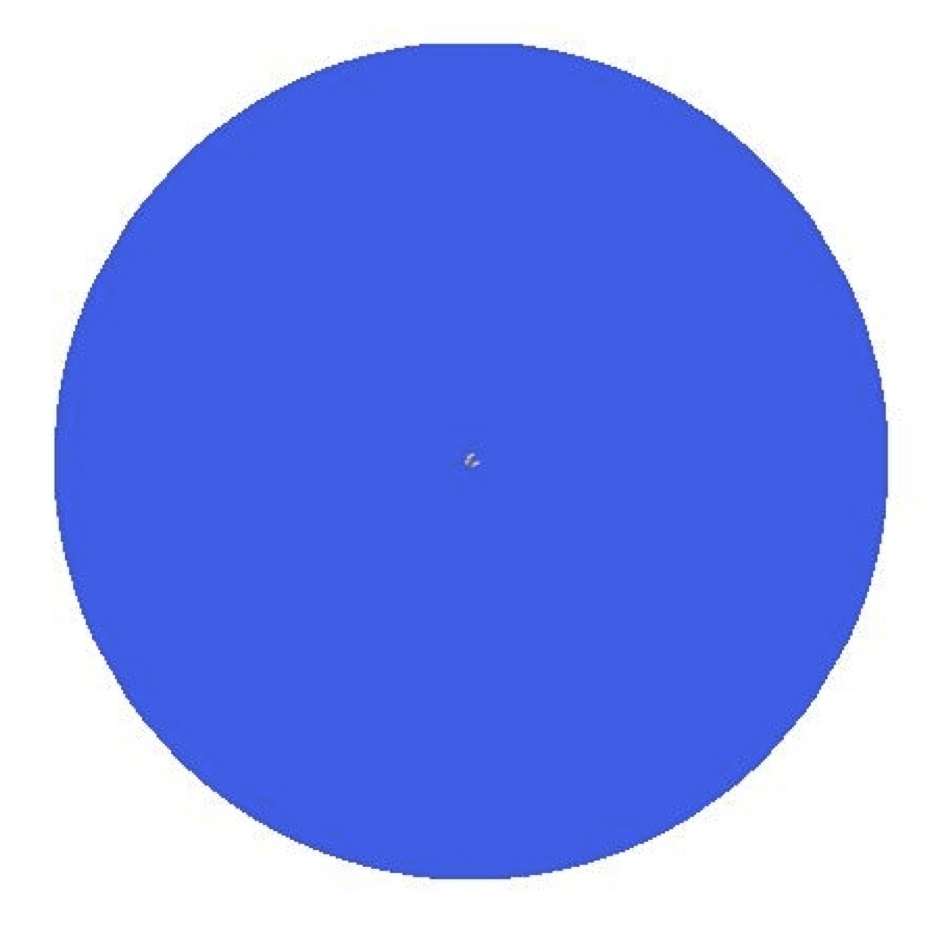}}&\resizebox{23mm}{!}{\includegraphics{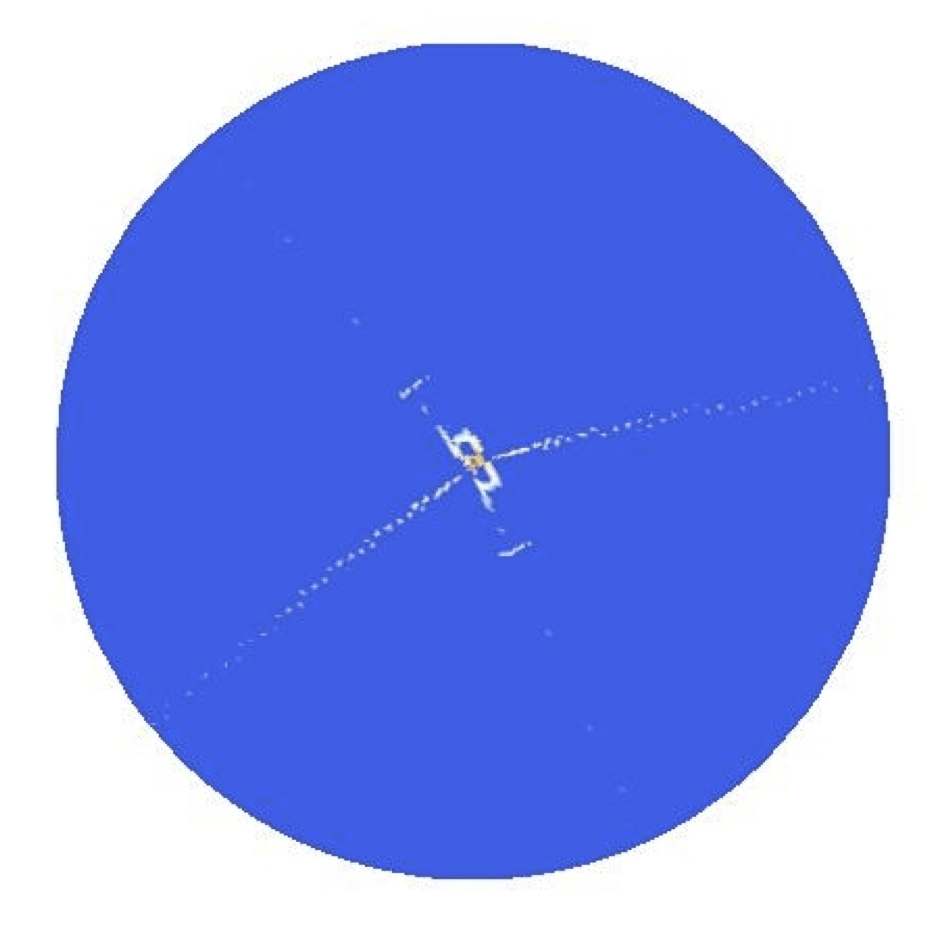}}&
\resizebox{23mm}{!}{\includegraphics{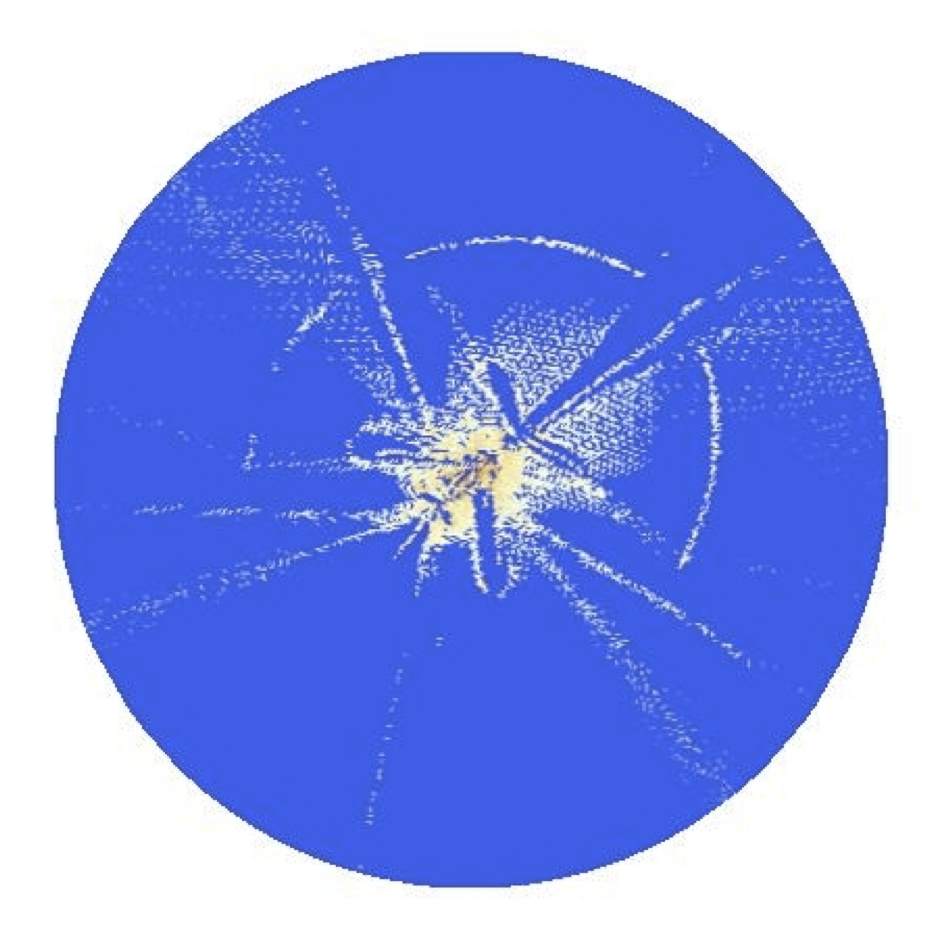}}&\resizebox{23mm}{!}{\includegraphics{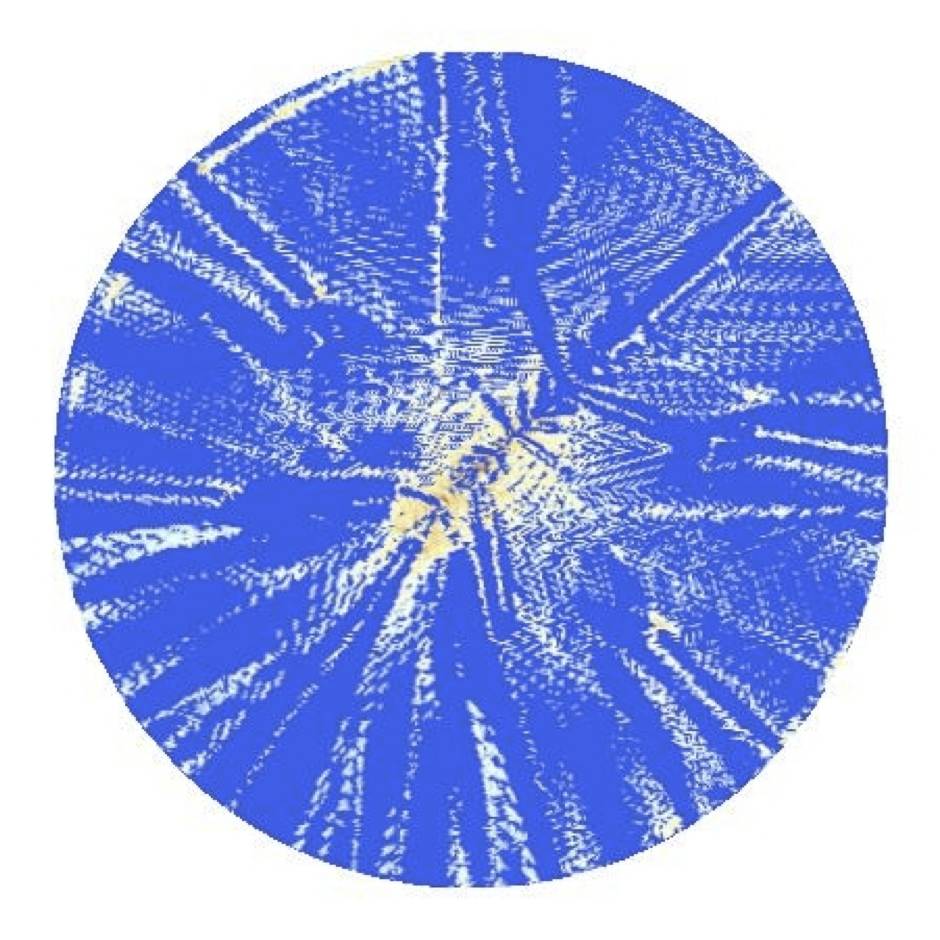}}
& \resizebox{23mm}{!}{\includegraphics{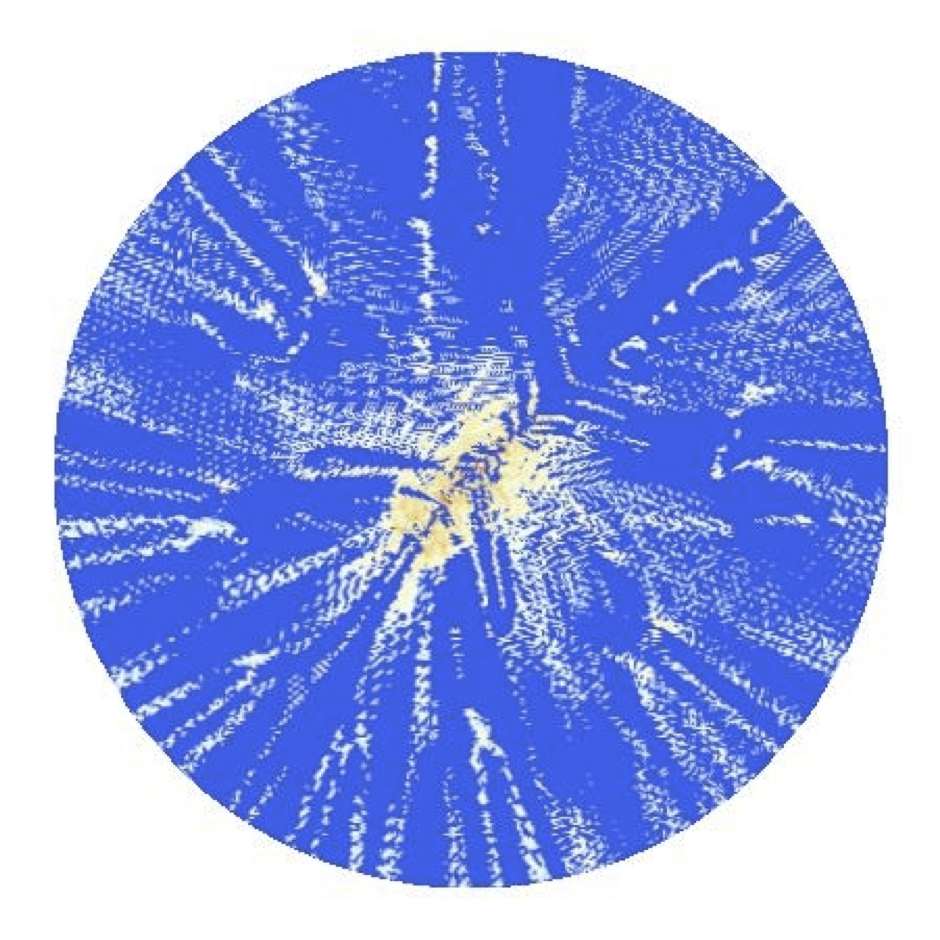}} \\

    \end{tabular}
\caption{\label{F5}(Color  online)  Configurations   from  molecular   dynamics
simulation, for $R_h=25\sigma$,  $R=290.7\sigma$, $k=5.7 \times 10^5
\epsilon/\sigma^2$ and  $J=2.85 \times 10^3\epsilon$.  The  (a.i) plots $(i=1,6)$ show the  3-D shape at
different packing stages: (1), The  appearance of the first d-Cone. (2),
The  first touch  of  the two  branches  of the  d-Cone. (3), The  two
branches start  to slide  over each other. (4), A highly  packed state
with  many folds. (5), The  appearance  of the  satellite d-cone. (6), Before the membrane escapes the  confining hole.  The (b.i) plots show
the cross-section of the membrane  in the plane of the confining hole.
Figures  (c,i),  (d.i),  and  (e.i),  Are  the  density  plots  of  the
stretching,  bending, and self-avoidance (excluded volume) 
energy, respectively,
projected back to the plane of the undeformed membrane.}
\end{center}
\end{figure*}

To understand the compaction of the membrane in more detail, we modeled 
a membrane dragged through a small hole using molecular dynamics (MD) 
simulations. This allows us to explore and analyze not only the mechanical 
response of the membrane, but also its energetics and compacted geometry.

We used  the computational  model of an  elastic sheet  constructed by
Seung  and  Nelson  \cite{seung1988}.  The  sheet  was  modeled  as  a
triangular lattice  with neighboring vertices bonded  by springs.  The
harmonic potential of each bond is
\begin{equation}
\label{BondPotential}
U_{\rm bond}(r)=\frac{1}{2}k(r-d)^2,
\end{equation}
where $k$ is the spring  constant and $d$ the equilibrium bond length.
We  express all  simulation parameters  and  results in  terms of 
the Lennard-Jones  (LJ) units based on the  LJ potential defined
as
\begin{equation}
\label{LJPotential}
V_{\rm   LJ}(r)=   4\epsilon_{\rm  LJ}\left[   \left(\frac{\sigma_{\rm
LJ}}{r}\right)^{12}-\left(\frac{\sigma_{\rm          LJ}}{r}\right)^{6}
\right],
\end{equation}
where $\epsilon_{\rm LJ}$ and $\sigma_{\rm LJ}$ are the characteristic
bending energy and length.  Their  units are denoted as $\epsilon$ and
$\sigma$, respectively.   LJ units are based  on $\epsilon$, $\sigma$,
and the  mass $m$ of the vertices.   For example, the unit  of time is
$\tau=\sqrt{m\sigma^2/\epsilon}$.    For   the   bond   potential   in
Eq.~\ref{BondPotential}  we used $k=5.7  \times 10^5\epsilon/\sigma^2$
and $d=2^{1/6}\sigma$.  The bending energy of the sheet is given by
\begin{equation}
U_{\rm bend}(\phi)=J (1+{\rm cos}\phi),
\end{equation}
where $J$  is the characteristic energy  of bending and  $\phi$ is the
dihedral angle  between two adjacent  triangles that share a  bond. For 
simulations reported here, $J=2.85 \times  10^3\epsilon$.  From $k$ and $J$,  one can derive
the  stretching and  bending moduli  of  the sheet  by identifying the 
low   energy    limit   of    this   lattice
model\cite{seung1988}.   In   our   case   the  bending   modulus   is
$B=2.46\times 10^{3}\epsilon$ and  the 2-D Young's modulus $Y=3.29\times
10^{6} \epsilon/\sigma^{3}$. The equivalent sheet thickness is given by $t=\sqrt{8J/k}=0.2 \sigma$.

\begin{figure}
\includegraphics[scale=0.65]{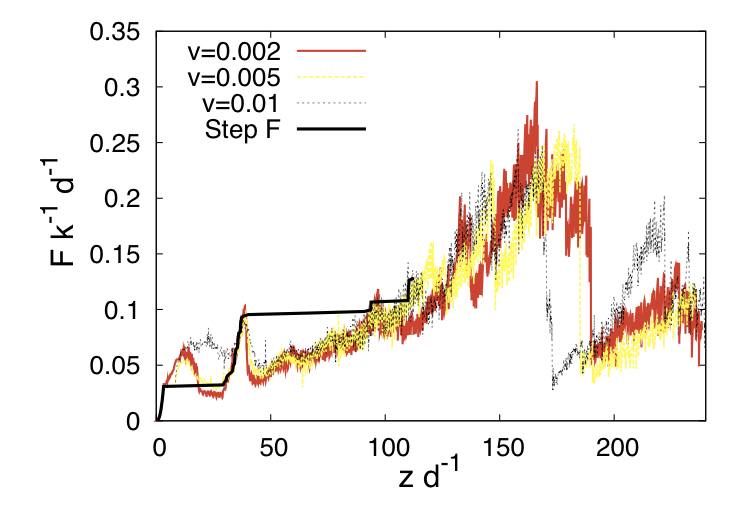}
\caption{\label{F6}(Color  online) Mechanical response of an in-silico membrane from MD simulations: Force
vs.  deformation curves  at  three different  pulling velocities, $v=0.002$ (continuos red line), $v=0.005$ (yellow dashed line) and $v=0.01$ (black dotted line), for
$R_h=25\sigma$, $R=290.7\sigma$, $k=5.7 \times 10^5 \epsilon/\sigma^2$
and  $J=2.85 \times  10^3\epsilon$. The solid (black) line is from a simulation 
where the dragging force is increased with small steps and the displacement 
of the membrane tip is measured. In this latter simulation MD is followed by 
energy minimization to allow the membrane to fully relax at any given dragging force.}
\end{figure}

The  self-avoidance  of  the  sheet  is imposed  by  adding  an  extra
interaction  between  all non-bonded  vertices.   This interaction  is
described by an LJ  potential with $\epsilon_{\rm LJ}=100\epsilon$ and
$\sigma_{\rm   LJ}=\sigma$.     The   potential   is    truncated   at
$r=2^{1/6}\sigma$ to make it  purely repulsive.  The same potential is
used to  impose the  interaction between the  sheet and  the confining
hole.   The later  is  implemented  as a  ring  of vertices  uniformly
distributed on a circle with radius $R_h$ and spacing $0.05d$.   

The  MD  simulations  were  performed using  the
Large-scale  Atomic$/$Molecular Massively Parallel  Simulator (LAMMPS)
developed at Sandia National Laboratories. A velocity-Verlet algorithm
with a  time step $dt=5\times  10^{-4}\tau$ was used to  integrate the
equations of motion. We performed two types of simulations.  In one, a
gradually increasing force was applied to the tip 
(the central vertex) of the sheet to pull it through the hole.  At a given
force,  MD  was run  for  a  certain  period (usually  $1000\tau  \sim
5000\tau$) and then an  energy minimization based on the Polak-Ribiere
version  of the conjugate  gradient algorithm  was performed.  In this
type of simulation, the energy of the sheet is minimized at each given
force. In other simulations, the sheet was dragged through the
hole by  pulling its tip at  a constant velocity
$v_z$ along  the vertical direction.  The resistance force on  the tip
from the  other part of  the sheet was  measured. We went to  very low
$v_z$ to reduce  the inertial effect. In this  case the measured force
was  found to  be  equal to  the force  exerted  on the  sheet by  the
confining hole within statistical errors. This indicate that the 
inertial effect is negligible.

The sheet is kept  at a very low temperature $T=10^{-4}\epsilon/k_{\rm
B}$ through a Langevin  thermostat. The Langevin damping rate $\Gamma$
is  $0.1\tau^{-1}$  for low  $v_{z}$.
For high $v_{z}$, higher damping rates have to be used. Tests with 
different $\Gamma$  showed that results reported here are not 
sensitive to the damping as long as the inertial effect is negligible.

\subsection{Numerical Results}
In  Fig.\ref{F5} we  summarize the  evolution  of the  geometry of  an
in-silico  membrane that is being  pulled through a hole.  The
results shown were obtained by  pulling the central vertex of the 
sheet at a constant speed. The configurations of the membrane 
shown in row $\bf{a}$ indicate that the model  does capture all
the regimes  found in  the experiments: The  formation of  the d-cone,
collision  of  the  main  central  fold,  Fig.\ref{F5} (a/2),  and  the
subsequent  formation  of  a  spiral-like structure with 
folds touching and sliding. Finally the  formation of
 the satellite  d-cone is  clear from
Figs.\ref{F5} ($a-b/5-6$).

To  qualitatively analyze  how  energy is
transferred  between   different  modes  (bending,   stretching,  self-avoidance)  
we  also  show  density  plots  of  the  different  energy
densities. It is clear  from Fig.\ref{F5} (c/4 and d/4)
 and Fig.\ref{F5} (c/5 and d/5),
that before the satellite d-cone is formed stretching concentrates
around the tip  and at the confining ring.  After the satellite d-cone
is  formed, stretching  goes down  and  the membrane  folds. Energy  is
transferred from the stretching mode  to the bending mode (see bright
scars  in  Fig.\ref{F5}  (d/5). 
The results in Fig. \ref{F5} indicate that the distributions 
of the bending and stretching energies are very heterogeneous 
and they are strongly localized near the tip of the 
membrane, the apex of the satellite d-cone, and the 
ridges along which the membrane is bent when 
the packing ratio is very high.

Overall, the mechanical responses of the in-silico membrane shown in Fig.\ref{F6} are 
qualitatively similar to those observed in our experiments. The membrane initially 
deforms through stretching but the conical symmetry is preserved. The conical 
symmetry breaks down when the d-cone appears, though there is 
still a  translational symmetry along the conical generators, at 
least outside the core of the d-cone. A slight drop in the dragging 
force accompanies the formation of the d-cone. Then the front of 
the d-cone moves inward and its two branches approach and collide. 
After this collision, the front of the d-cone keeps moving towards and 
finally collides with the opposite side of the membrane. After this 
point the two contacting branches of the d-cone start to flatten in 
response to the increasing confinement. At certain dragging force the 
two branches start to slide, inducing a clear drop in the measured 
dragging force on the membrane. This corresponds to the first major bump 
in the $F$-$z$ curve. When the confinement is further increased, more branches 
and folds are born, collide, and slide over each other. Each sliding event 
accompanies a drop in F, appearing as a bump in the $F$-$z$ curve. But overall 
F increases as the deformation grows. It should be emphasized that the 
translational symmetry along the conical generators is still preserved in this stage.

When the displacement of the membrane tip exceeds a certain amount, or 
equivalently when the confinement is strong enough, the translational 
symmetry breaks down with the appearance of a satellite d-cone. The apex 
of this second d-cone is away from the membrane tip. It appears in the form 
where many folds buckle and bend towards the interior of the compacted membrane. 
At this catastrophic event, the dragging force F drops dramatically by an order 
of magnitude, shown as a steep drop in the $F$-$z$ curve. After this event, 
$F$ starts to increase again as the confinement is further increased. 

\begin{figure}
\includegraphics[scale=0.65]{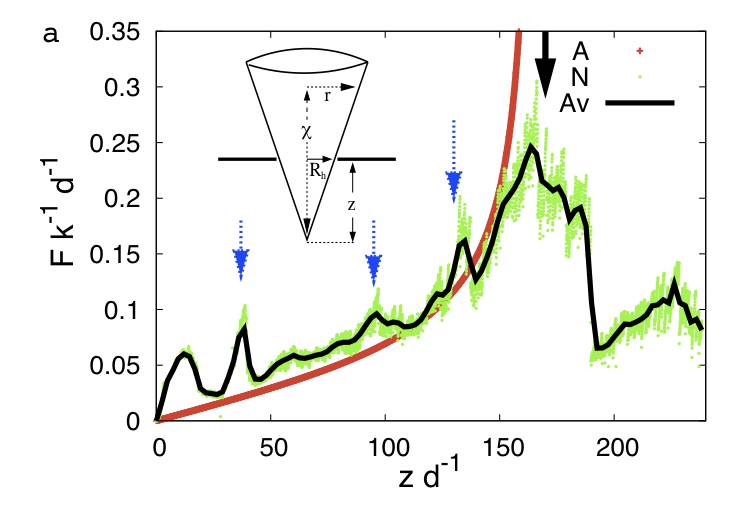} 
\includegraphics[scale=0.65]{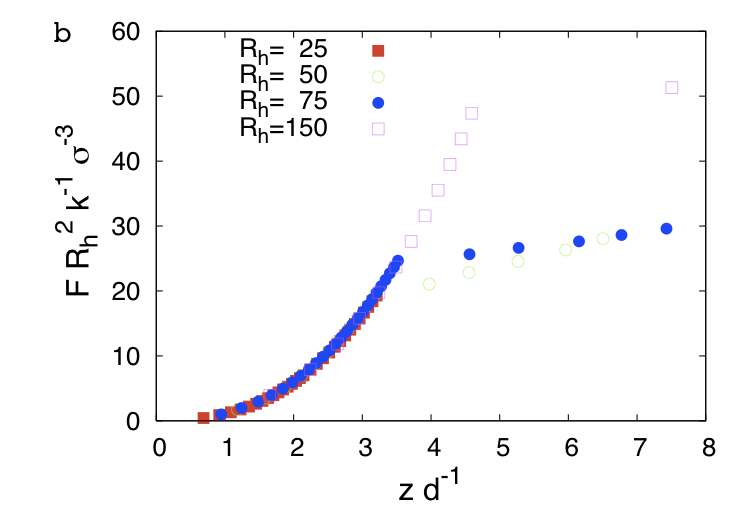}
\caption{\label{F7}(Color  online) Mechanical response of the
membrane   for  increasing   confinement: (a),  Points (green)
correspond to one set of numerical data (N), showing many stick-slip events
(from the left first, second and third arrows) as  well  as  a  huge  relaxation  for $z\sim  175 d$ denoted by the rightmost (black) arrow.  The  darker solid (black)  line
is from the  averaged data (Av) and the monotone (red) line (A) is the  best fit
obtained from the model described in the text. The inset shows a schematic representation of  the geometry used for the analytic calculation of the mechanical response at moderate packing, where $z$ is the vertical distance from the tip, $R_{h}$ is the radius of the hole and $r$ is the polar radius of the cone at a vertical distance $\chi$ from the tip. (b), In the low load regime the
$F-z$ curves collapse into  a single universal function $f(z)$ defined
by Eq.(\ref{scaling1}).  $R_{h}$ denotes radius of the hole.}
\end{figure}

Though quantitative agreement is not expected, there is also an obvious qualitative difference 
in the mechanical responses of the membrane between numerical and experimental 
results. For example,  in the second regime after the birth of the d-cone and 
the appearance of the satellite d-cone, more bumps in $F$ are observed in the numerical 
results (three  bumps marked  with  three  blue
arrows  in the  simulations of Fig.\ref{F7}a,  instead of  the one  observed in
experiments). These bumps correspond to the consecutive collision and sliding of folds. The  reason for
this  difference is that in-silico  membranes   are   softer   and   less
frictional. Fig.\ref{F3} shows that  the  less
frictional  membrane, the  one  lubricated with  graphite, shows  more
bumps than  the others, closely resembling the  numerical results.
However, it should be emphasized that due  to the roughness  of the 
iso-potential lines that describe  the \emph{surface} of 
the in-silico  membrane, there is
an  effective friction,  even  though  we did  not  put any  explicit
friction law in  the simulation. The effective friction  is an analog
of the tangential force that would appear when we try to slide two egg
boxes that  are on top of  each other. Each atomic  stick-slip process
appears as  a sudden bump in the  $F$-$z$ curve. 
This is also the reason that the numerical results are noisy, albeit 
quite reproducible for each dragging velocity.

Several experiments on crumpling were performed by compressing sheets 
with increasingly heavier weights \cite{matan2002}. 
In our set-up, this corresponds to applying a dragging force on the membrane 
that increases gradually. Results showed that this approach gives results 
different from those from experiments where the confinement or deformation 
of the sheet is controlled. It should be emphasized that experiments performed 
with increasing forces do not provide detailed information about the relaxation 
of the membrane, while such information is available from experiments performed 
with deformation controlled and mechanical response measured.

Simulations also allow us to vary the dragging force $F$ and measure the vertical 
displacement $z$ of the membrane tip. The latter is one signature of the mechanical 
response of the membrane. The solid (black) curve in Fig.\ref{F6} shows one $F$-$z$ curve 
generated in such a simulation. In the simulation the dragging force is increased 
in a stepwise manner. In most cases, the membrane reached a steady crumpled state 
quickly and $z$ was easily obtained. However, at certain values of $F$, the membranes 
took a long time to relax. Typically, each relaxation corresponds to the transition 
from one locally stable compact configuration to another. For example, when the 
load reached the critical value at which the d-Cone appeared, and was kept at this 
point, the fold grew and moved towards the opposite side of the membrane, and its 
two branches approached each other. This process ends when the two branches touched 
and the front of the fold contacted the opposite side of the membrane. The membrane 
entered a new configuration, stabilized by the effective friction. This transition corresponds to the first 
plateau in the $F$-$z$ curve as shown in Fig.\ref{F6}. Successive plateaus 
were observed at the occurrence of strong reorganization of folds, including 
the birth of new folds or abrupt interfacial sliding of contacting 
folds. The results here make it clear that the full $F$-$z$ curve can 
only be obtained when $z$, not $F$, is controlled. The underlying reason is 
simply that $z$ is not a single-valued function of $F$ due to the 
hardening-softening of the membrane.

\subsection{D-cone formation}
In real experiments  it is extremely difficult to  measure the initial
deformation  of the  membrane. Intuition  tells us  that for  any real
membrane  at low  loads it  should stretch,  and then,  once  the stored
energy  is sufficient  to produce  a global  change, a  d-cone  will be
formed. This regime is readily explored for in-silico membranes. Results show that 
the $F$-$z$ curve grows in a quasi-parabolic fashion, $F\sim z^{1.33 \pm
0.02} $. This  indicates that pure bending  is not the relevant
mode at extremely low loads. Otherwise, $F\sim z$ as predicted for a
perfectly inextensible membrane \cite{cerda1999}.

In Fig.\ref{F7}b, the $F$-$z$ curve at the early stage of
the folding process  is shown. Here numerical data  for four different
$R_{h}$  are plotted. Results show that the force has a universal functional form given by
\begin{eqnarray}
F=f(z)\frac{k \sigma^{3}}{R_{h}^{2}}
\label{scaling1}
\end{eqnarray} 
This is  a clear  signature that at  extremely low loads  the membrane
first stretches, and the d-cone only appears when a critical load $F_{c}$ is reached at $z\sim 5d$. After the occurrence of the d-cone, the main  energy contribution
comes  from the  bending of  the membrane,  and the  scaling  shown in
Eq.\ref{scaling1}  breaks  down.  This is seen from 
the discontinuity of the mechanical
response at $z\sim 5d$ that marks  the onset  of  the first  d-cone.  The  correct
scaling for  the force just after  the first d-cone is  formed is well
known, $F\sim  z B/R_{h}^{2}\log(R/r_{c})$,  where $B$ is  the bending
stiffness  and $r_{c}=5$ mm  is the  core radius which is of the order of the crescent singularity.   This scaling  works for
small deformations (small confinement) $z/R_{h}<1/4$ \cite{cerda1999}.

\subsection{Mechanical response at moderate packing.}
We  used a  phenomenological  route to  explain  $F$-$z$ at  moderate
confinement after the first d-cone  is formed and before the emergence
of the  satellite d-cone ( $5d< z < 175d $ in  Fig.\ref{F7}a)
in terms of the bending energy  of the system only. A similar approach
 was used by Boue \textit{et al.} \cite{boue2006}.  

We consider the bending energy in this
regime as  the sum  of the  bending energy of  two regions  that fold,
qualitatively, in  a different way (look at the inset of Fig.\ref{F7}a for clarity). The
first contribution to  the bending energy is from the isolated big
fold living in the unoccupied core region.   The second region contributing to the bending energy is the sector of the membrane far from the center of 
the cone. 

The length of the enclosed inextensible membrane at the hole is:
\begin{eqnarray}
L&\sim& 2 \pi \sqrt{R_{h}^2+z^2}
\label{e1}
\end{eqnarray}

$L$ determines the maximum polar angle:
\begin{eqnarray}
\theta_{\rm{max}}&\sim& 2 \pi \sqrt{\frac{z^2}{R_{h}^2}+1}
\label{e2}
\end{eqnarray}
Since the bending energy  of the core region  scales with the
square of its curvature, we have that:
\begin{eqnarray}
U_{\rm{bend}}^{\rm{core}}                                                &\sim&
\int_{\mathcal{C}}\int_{z}R_{\rm{core}}^{-2}ds dz
\label{e3}
\end{eqnarray}
The characteristic radius for the fold living in the core
region can be estimated as:
\begin{eqnarray}
R_{\rm{core}}&\sim&\sqrt{ r^2-2 t \sqrt{r^2+\chi^2}}
\label{e4}
\end{eqnarray}
where $r$ is  the polar radius of the cone at  a vertical distance $\chi$
from the tip and $t$ is the thickness  of the membrane.  Placing this back
into Eq.\ref{e3} we  get the final form for the  bending energy of the
core:
\begin{eqnarray}
U_{\rm{bend}}^{\rm{core}}\sim      \pi B \xi \sqrt{(1+\xi^{-2})(1+\xi^{2})} \times \nonumber \\
\log{\left(1+\frac{(R-r_{c})}{r_{c}-t(1+\xi^{2})}\right)}
\label{e5}
\end{eqnarray}
where R is the radius of the membrane and $\xi=z/R_{h}$. In the limit when $t \rightarrow 0$, the logarithmic contribution approaches $\log{(R/r_{c})}$. This simple model yields the signature of a divergent response when $r_{c}=t(1+\xi^{2})$, evidencing  that there is a critical distance $z$ at which becomes extremely expensive to deform the core region. The critical pulling distance is given by: $z_{c}=R_{h}\sqrt{r_{c}/t - 1} \sim 165 d$ which compares well with the numerical data ($z=175 d$ see Fig.\ref{F7}a, black arrow).

The second region is illustrated by the cross-section cut of the 
membrane shown in Fig.\ref{F2}. We model this region as an archimedian spiral. Its radius is 
\begin{eqnarray}
R_{\rm{spiral}}&\sim&\sqrt{ r^2-2 t \sqrt{r^2+\chi^2}}+\frac{t\theta}{2\pi}
\label{e6}
\end{eqnarray}
The bending energy from this spiral region is:
\begin{eqnarray}
U_{\rm{bend}}^{\rm{spiral}}\sim   \pi^{1/2}B \xi \sqrt{(1+\xi^{-2})(1+\xi^{2})} \times \nonumber \\
\log{\left(\frac{(1+R/t)+\xi^{2}}{(1+r_{c}/t)+\xi^{2}}\right)}
\label{e7}
\end{eqnarray}

\begin{figure}
\includegraphics[scale=0.65]{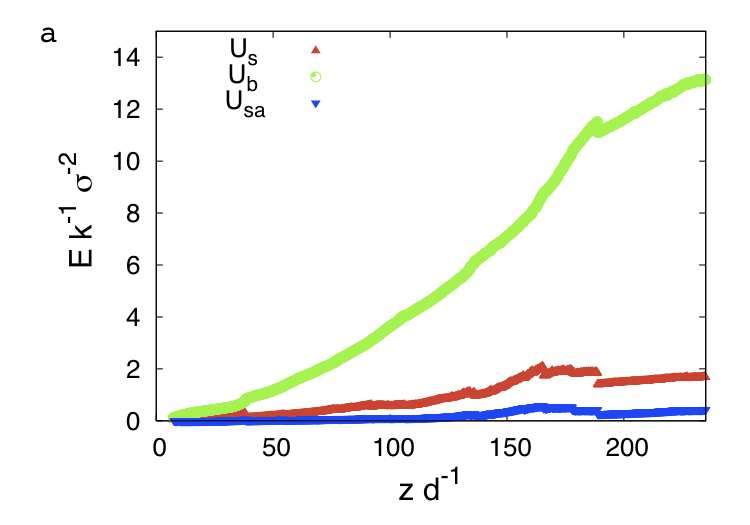}
\includegraphics[scale=0.65]{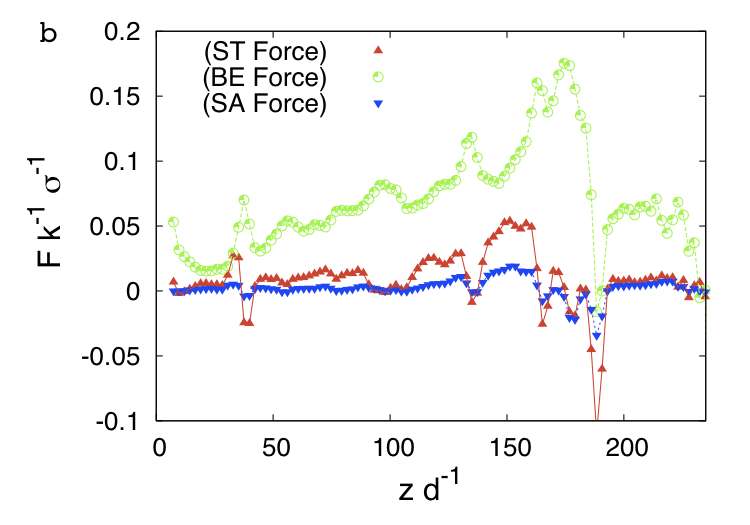}
\caption{\label{F10}(Color  online)  Development  of  different   interactions  for
increasing confinement.  (a), Stretching (middle curve, red triangles pointing up), bending (top curve, green circles),
and   self-avoidance  (bottom curve, blue triangles pointing down)   energy  as   a  function   of  vertical
displacement. (b), Individual  contributions to  the total force   arising    from     bending (top, green circles),    stretching (middle, red triangles pointing up),    and    self-avoidance (bottom, blue triangles pointing down) interactions. 
Transitions between  different regimes at $z \sim  5 d$ and
$z \sim  185 d$  indicate the  formation of the  first d-cone  and the
satellite d-cone.}
\end{figure}

Finally, the total energy (approximated as the bending energy only) at moderate packing, $U_{\rm{total}}^{\rm{analytic}}\sim \alpha U_{\rm{bend}}^{\rm{core}}+U_{\rm{bend}}^{\rm{spiral}}$ where $\alpha$ is a numerical constant to be determined. Estimating $r_{c}=7d$ from the geometry of the deformed sheet from the numerical simulation (the size of the crescent in Fig.\ref{F5}) and considering the constant $\alpha$ and the bending stiffness $B$ as a free parameters in Eqs.\ref{e5} and \ref{e7}, we have performed a numerical best fit of our analytic expression $U_{\rm{total}}^{\rm{analytic}}$ and compared it with the numerical bending energy shown in the top curve of Fig.\ref{F10}a. From this operation we obtained $B=2.463\times 10^{3} \epsilon$ and $\alpha = 3.26$. Using these values in $U_{\rm{total}}^{\rm{analytic}}$, we computed the total force in this regime, $F_{\rm{total}}^{\rm{analytic}}$, which is shown in the monotone (red) curve in Fig.\ref{F7}a. Even though we have completely
neglected self-avoidance and stretching, we obtained qualitative agreement in the mechanical response between our analytic approach and the numeric results, up to the point at which a satellite d-cone develops.
This agreement indicates that in this regime the deformation of the membrane 
is dominated by the bending mode. This observation is further strengthened 
by decomposing the total energy into bending, stretching, and self-avoidance 
contributions, Fig.\ref{F10}a, which shows that bending energy is larger than the other 
two by an order of magnitude. 

\subsection{First and second generation d-cones} 
The catastrophic relaxation  event in the $F$-$z$ curve shown by the rightmost (black) arrow in Fig.\ref{F7}a  is the  signature of the
birth  of  a second  generation  d-cone in  the  real  and in-silico
membranes.  The mechanical response  of the
system after this new singularity was born can be rationalized in terms
of a renormalized bending stiffness of the sheet which now resembles a
cone that is thicker than the original membrane. Indeed, since bending
energy     of   a  d-cone     scales     as     $U_{\rm{bend}}\sim     Bz^{2}/R_{h}^{2}
\log\left(R/r_{c}\right)$ for  small deformations, we  expect that the
ratio between the  response of the first d-cone  and the second d-cone
will  be  proportional  to  $(t_{eff}/t)^{3}$, where $t_{eff}=nt$ and
$n$  is  the number  of  folded  layers,  apart  from  geometric
factors due to the change in the contact angle.   From numerical  simulations
results shown in Fig.\ref{F5} we  see that $n\sim 4$ at the onset of
the second d-cone.  By comparing, on the other hand,  the slope of the
numeric   $U_{B}$-$z$  curve  (Fig.\ref{F10}a   top  curve)   we  get
$n\sim(\frac{B_{2}}{B_{1}})^{1/3}\sim 3.2$. The agreement 
is reasonable,
considering that  the four  layers are not  glued together. Note
that membranes  with very small  coefficient of friction  will deviate
from this  simple estimate, since in  that case the  folded layers are
not going  to show a cooperative  behavior but most  likely will slide
respect to each other.

Using our argument of the  renormalized stiffness constant, we turn to
compute the  number of  folded layers for  which the  satellite d-cone
would  be created. Examining  Fig.\ref{F10}b we  see that  bending and
stretching  forces   become  comparable  at  the  onset   of  the  new
d-cone. The reason  is that once the membrane  has become thick enough
and is stuck due to  the geometrical constraint, it has to stretch if 
the dragging force $F$ is further increased. The satellite d-cone only 
occurs when the stretching component of the force becomes comparable 
with the bending component. Matching these two  forces yields  $n=3$, in
fair agreement with the in-silico value.

\section{Concluding Remarks}  
We have  shown that under generic conditions,  the mechanical response
of an elastic membrane exhibits a singular evolution. This response is
a  consequence  of  bending  minimization,  the  frictional  collision
between folds, and the birth of conical singularities.

In particular, we identified the satellite d-cone as a basic deformation 
mode of a moderately folded membrane when further constrained, which can 
be effectively described as a conical membrane with a larger thickness 
determined by the number of the folded layers. This is further supported 
by our experiments starting with a single-layer conical membrane. 
Our preliminary results and theoretical analysis show that the location 
of the tip of the satellite d-cone depends on the opening angle of the cone.

Based on  our experimental  observations and numerical  simulations, we 
suggest  a new  route  to  model  the mechanical  response  of
crumpled surfaces,  by considering an effective  system that preserves
the overall symmetries of the deformed one and that has a renormalized
bending  stiffness.  It seems feasible to use this approach to expedite 
numerical simulations of crumpling and to make easier theoretical analysis of crumpled surfaces.

We also showed that for any  real and in-silico membranes the stretching mode dominates
until the stretching energy is
enough to produce a global change, leading to the birth of the d-cone. Before the
formation of  the d-cone we showed  that there is  a universal scaling
function for the dragging  force that  shows   an   anomalous  exponent   in  the   vertical
displacement.

Finally  we showed that  friction  does play  an important  role
in modifying the mechanical response of the membrane. Its effects emerge 
soon after the first collision event between the two branches of the d-cone.  
We  showed  that  friction will  generically
enhance the above mentioned renormalization  process. It is expected that, as friction is increased, a d-cone appears sooner. This was confirmed 
by our experiments on membranes coated with materials having different frictional properties. 

We expect our findings to  be applicable to enhance energy dissipation
in structures  under extreme deformations by  suitable modification of
its frictional properties. For  example, in car collisions more energy
could be dissipated by designing  the inner surface of metallic sheets
to be structured in such a way that as soon as metal plates touch each
other they get stuck. It would be interesting to see if this type of 
ideas can help us design structures with enhanced response to impacts.
\section*{Acknowledgments}
We  thank  O.  Tchernyshyov,  I.  Cabrera and  E.  Felton  for  useful
comments.   P.M.  acknowledges  support  from NSF  Grant  DMR-0520491.
S.C. acknowledges support from NSF Grant CMMI-0709187.

\end{document}